\documentclass[3p]{elsarticle}
\usepackage{lineno,hyperref}
\usepackage{graphicx}
\usepackage{amsmath}
\usepackage{color}
\usepackage[normalem]{ulem}

\journal{arXiv}
\newcommand\myFigureWidth{0.30}

\begin{document}

\begin{frontmatter}

\title{Distributions of Historic Market Data -- Stock Returns}

\author[mymainaddress]{Zhiyuan Liu}
\author[mymainaddress]{M. Dashti Moghaddam}
\author[mymainaddress]{R. A. Serota\fnref{myfootnote}}
\fntext[myfootnote]{serota@ucmail.uc.edu}

\address[mymainaddress]{Department of Physics, University of Cincinnati, Cincinnati, Ohio 45221-0011}

\begin{abstract}
We show that the moments of the distribution of historic stock returns are in excellent agreement with the Heston model and not with the multiplicative model, which predicts power-law tails of volatility and stock returns. We also show that the mean realized variance of returns is a linear function of the number of days over which the returns are calculated. The slope is determined by the mean value of the variance (squared volatility) in the mean-reverting stochastic volatility models, such as Heston and multiplicative, independent of stochasticity. The distribution function of stock returns, which rescales with the increase of the number of days of return, is obtained from the steady-state variance distribution function using the product distribution with the normal distribution.
\end{abstract}

\begin{keyword}
Stock Returns \sep Volatility \sep Heston \sep Multiplicative \sep Distribution Tails
\end{keyword}

\end{frontmatter}

\section{Introduction\label{Introduction}}
Historic data of stock returns (SR) and volatility can provide invaluable testing of stochastic models and approximations used to describe the markets. While the model for SR is mostly established, the models for stochastic volatility are numerous and there has not been conclusive evidence yet that decidedly favored one of them. In this paper we concentrate on two mean-reverting models of volatility: Heston \cite {heston1993closed,dragulescu2002probability} and multiplicative \cite{ma2014model}. The former model is characterized by the Gamma (Ga) steady-state distribution of variance (squared volatility) and short tails of the SR distribution; the latter is characterized by fat tails -- Inverse Gamma (IGa) for variance and Student's for SR. 

Unfortunately, neither direct fits of the SR distribution nor investigation of its tails is capable of determining the better model, as will be shown in the bulk of the paper. Fortunately, the moments of the SR distribution are not only decisively better fitted with the Heston model (HM), but also in excellent agreement with theoretical predictions. The latter are obtained by using the product distribution (PD) to find the distribution function of SR, which simply scales with the number of days over which SR are calculated. We also show that Ito-calculus corrections become important relative to the PD only over long times, when SR distribution would approach its normal distribution limit.

This paper is organized as follows. In Section \ref{ModelsRV} we define the models and obtain, analytically and numerically, the mean realized variance (RV) of SR as a function of the number of days of returns. In Section \ref{DistrSR} we derive the main analytical results for the distribution function of SR and discuss the Ito-calculus corrections and approach to normality. In Section \ref{Numerics} we perform numerical fitting of the SR distribution functions. In Section \ref{Moments} we compare the numerical results for the moments of the SR distribution function with analytical predictions.

\section{Models of Volatility and Realized Volatility\label{ModelsRV}}

The standard equation for the stock price reads
\begin{equation}
\mathrm{d}S_t = \mu S_t \mathrm{d}t + \sigma_t S_t \mathrm{d}W_t^{(1)}
\label{sdrSt}
\end{equation}
where $\sigma_t$ is the  volatility. Denoting variance $v_t=\sigma_t^2$,  $r_t = \ln (S_t/S_0)$ and $x_t = r_t - \mu t$ and using Ito calculus, the equation for log returns becomes
\begin{equation}
\mathrm{d}x_t = -\frac{\sigma_t^2}{2}\mathrm{d}t + \sigma_t\mathrm{d}W_t^{(1)} =- \frac{v_t}{2}\mathrm{d}t + \sqrt{v_t}\mathrm{d}W_t^{(1)}
\label{sdrxt}
\end{equation}
The two models of stochastic volatility studied here are Heston \cite{dragulescu2002probability}
\begin{equation}
\mathrm{d}v_t = -\gamma(v_t - \theta)\mathrm{d}t + \kappa \sqrt{v_t}\mathrm{d}W_t^{(2)}
\label{sdrGaGa}
\end{equation}
and multiplicative \cite{ma2014model}
\begin{equation}
\mathrm{d}v_t = -\gamma(v_t - \theta)\mathrm{d}t + \kappa v_t\mathrm{d}W_t^{(2)}
\label{sdrIGaIGa}
\end{equation}
Both $\mathrm{d}W_t^{(1)}$ and $\mathrm{d}W_t^{(2)}$ are distributed normally, $\mathrm{d}W_t^{(1)} \sim \mathrm{N(}0,\, \mathrm{d}t \mathrm{)}$, $\mathrm{d}W_t^{(2)} \sim \mathrm{N(}0,\, \mathrm{d}t \mathrm{)}$.

It should be pointed out that in general $\mathrm{d}W_t^{(1)}$ and $\mathrm{d}W_t^{(2)}$ are correlated as
\begin{equation}
\mathrm{d}W_t^{(2)} = \rho \mathrm{d}W_t^{(1)} + \sqrt{1-\rho^2} \mathrm{d}Z_t
\end{equation}
Where $\mathrm{d}Z_t$ is independent of $\mathrm{d}W_t^{(1)}$, and $\rho \in [-1,\, 1]$ is the correlation coefficient. The latter can be evaluated from leverage correlations \cite{bouchaud2001leverage,perello2002correlated,perello2004multiple} and its calculated magnitude is very model-dependent. However, since $x_t$ does not enter into the equation for volatility (variance) directly, we surmise that the steady-state distribution of the latter does not depend on $\rho$. Further, it will be argued and confirmed below that the distribution of the SR can be found as a PD of the afore-mentioned $\rho$-independent steady-state distribution of stochastic volatility and normal distribution. Consequently, for the purposes of this paper, we can set $\rho=0$. (This was empirically observed for HM \cite{dragulescu2002probability} -- see the comment below in the paragraph following (\ref{pdfJM}).) Furthermore, since it will be also argued below that the first term in the r.h.s. of (\ref{sdrxt}) does not yield significant corrections to the SR distribution until very long periods of returns, it can be neglected and (\ref{sdrxt}) can be rewritten as
\begin{equation}
\mathrm{d}x_t \approx \sigma_t\mathrm{d}W_t^{(1)}
\label{sdrxtapprox}
\end{equation}
One way to mathematically interpret this approximation is to realize that ``small'' $\mathrm{d}t$ can actually be very long in terms of SR.

\begin{figure}[!htbp]
\centering
\begin{tabular}{cc}
\includegraphics[width = \myFigureWidth \textwidth]{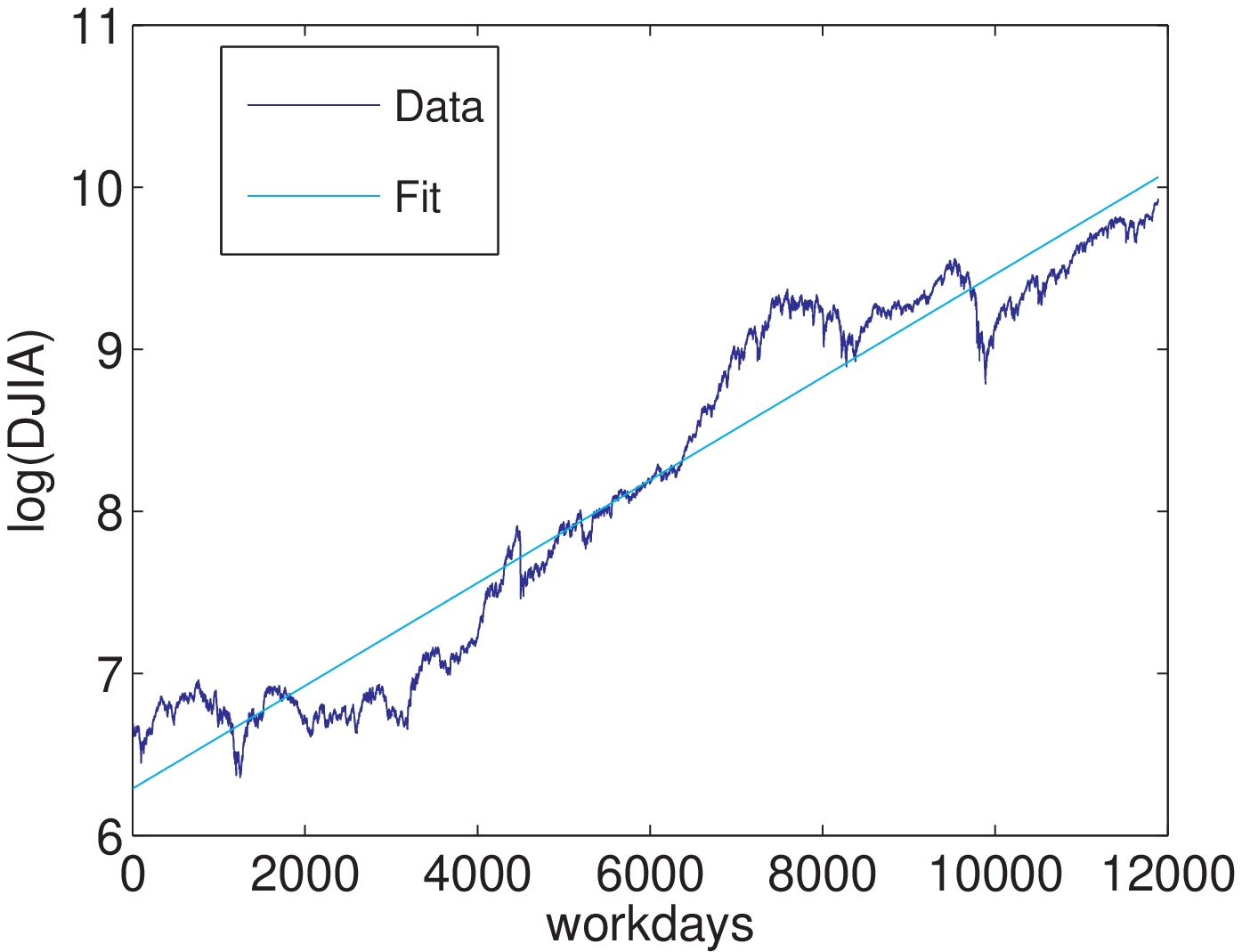}
\includegraphics[width = 0.32 \textwidth]{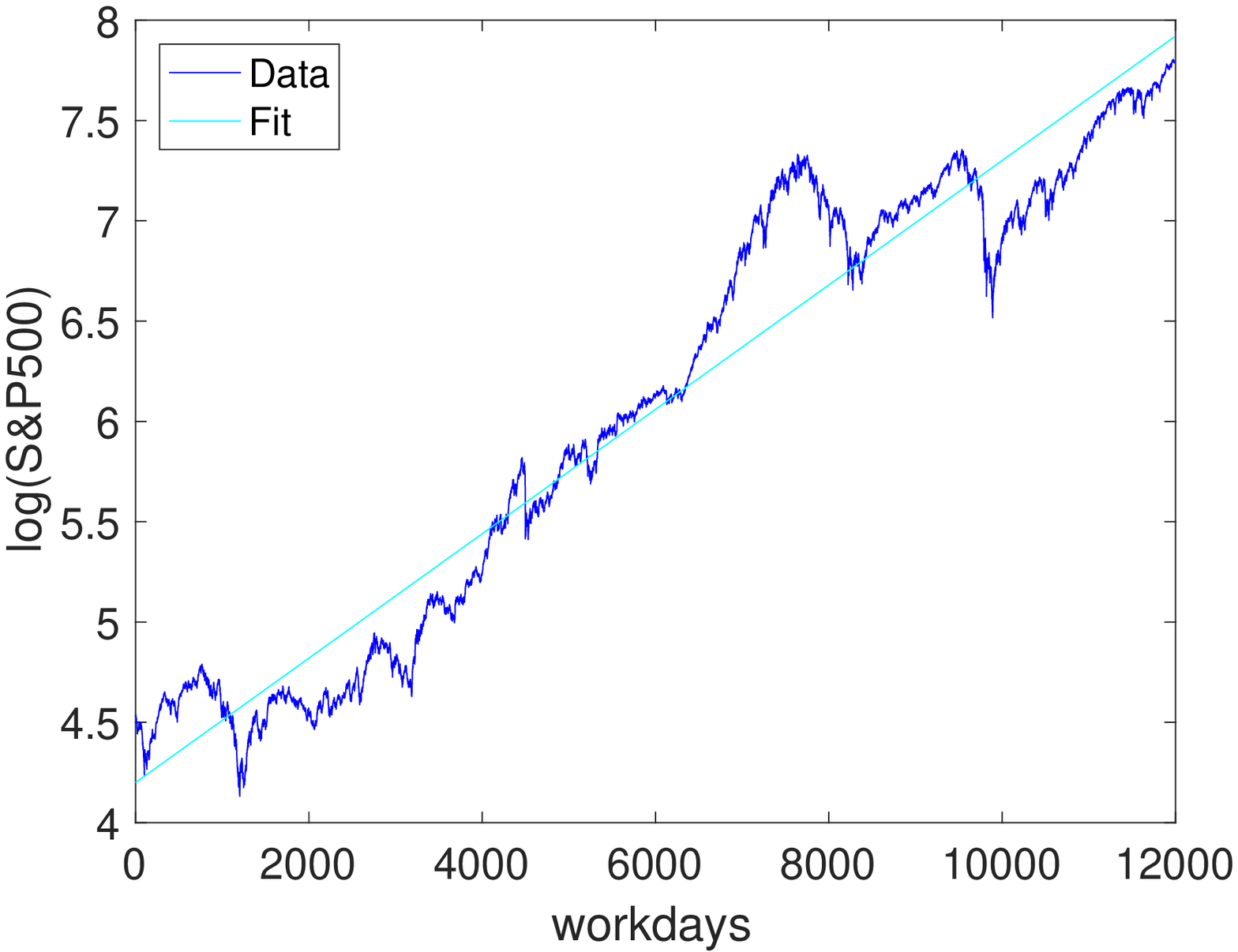}\\
\end{tabular}
\caption{Historic data: DJIA (left) and S\&P500 (right). The best fit growth rate are $\mu_{\text {DJIA}} = 3.17 \times 10^{-4}$ and $\mu_{\text {S\&P}}=3.10 \times 10^{-4}$.}
\label{GoogleHistoricData}
\end{figure}

\begin{figure}[!htbp]
\centering
\begin{tabular}{cc}
\includegraphics[width = 0.33\textwidth]{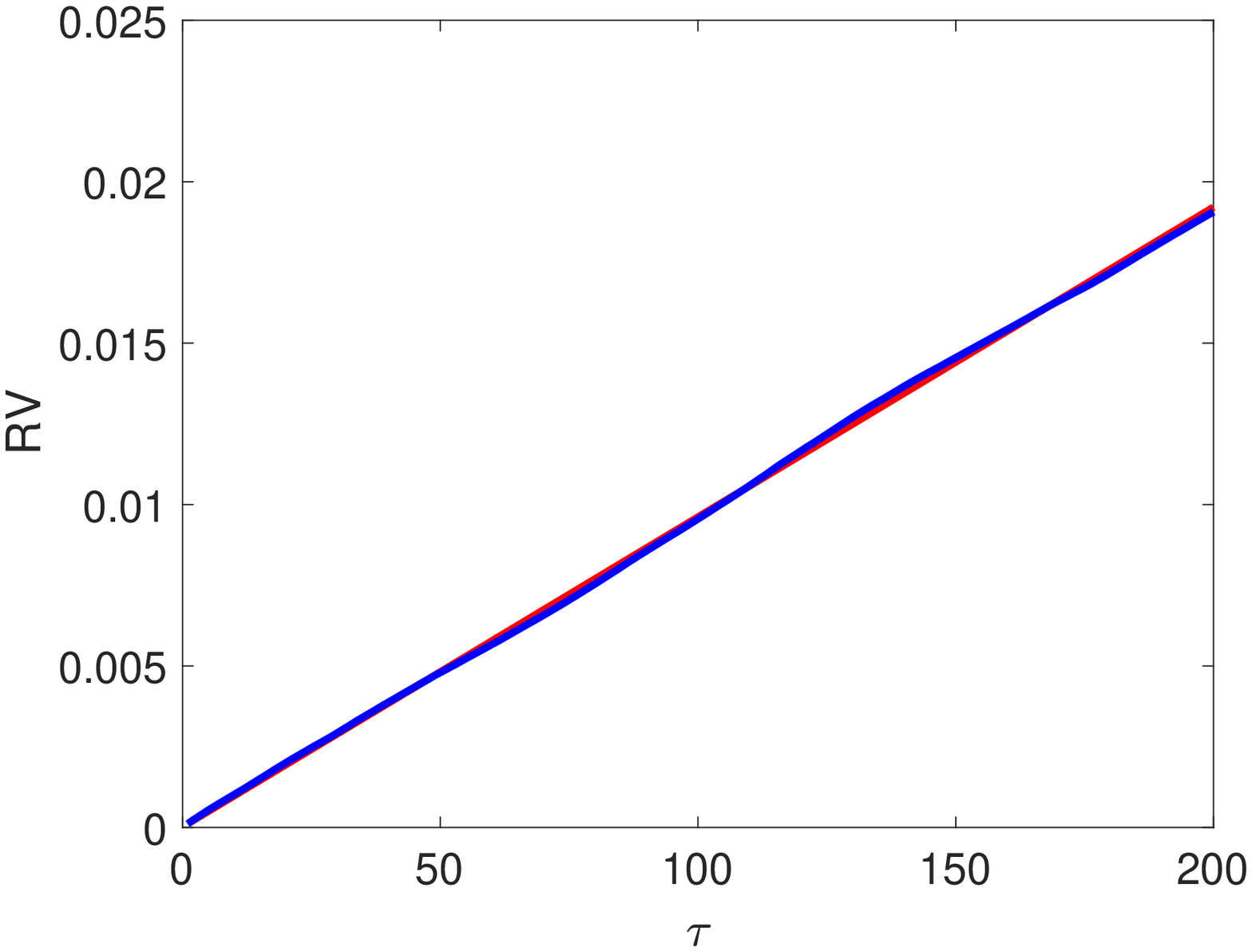} &
\includegraphics[width = 0.33\textwidth]{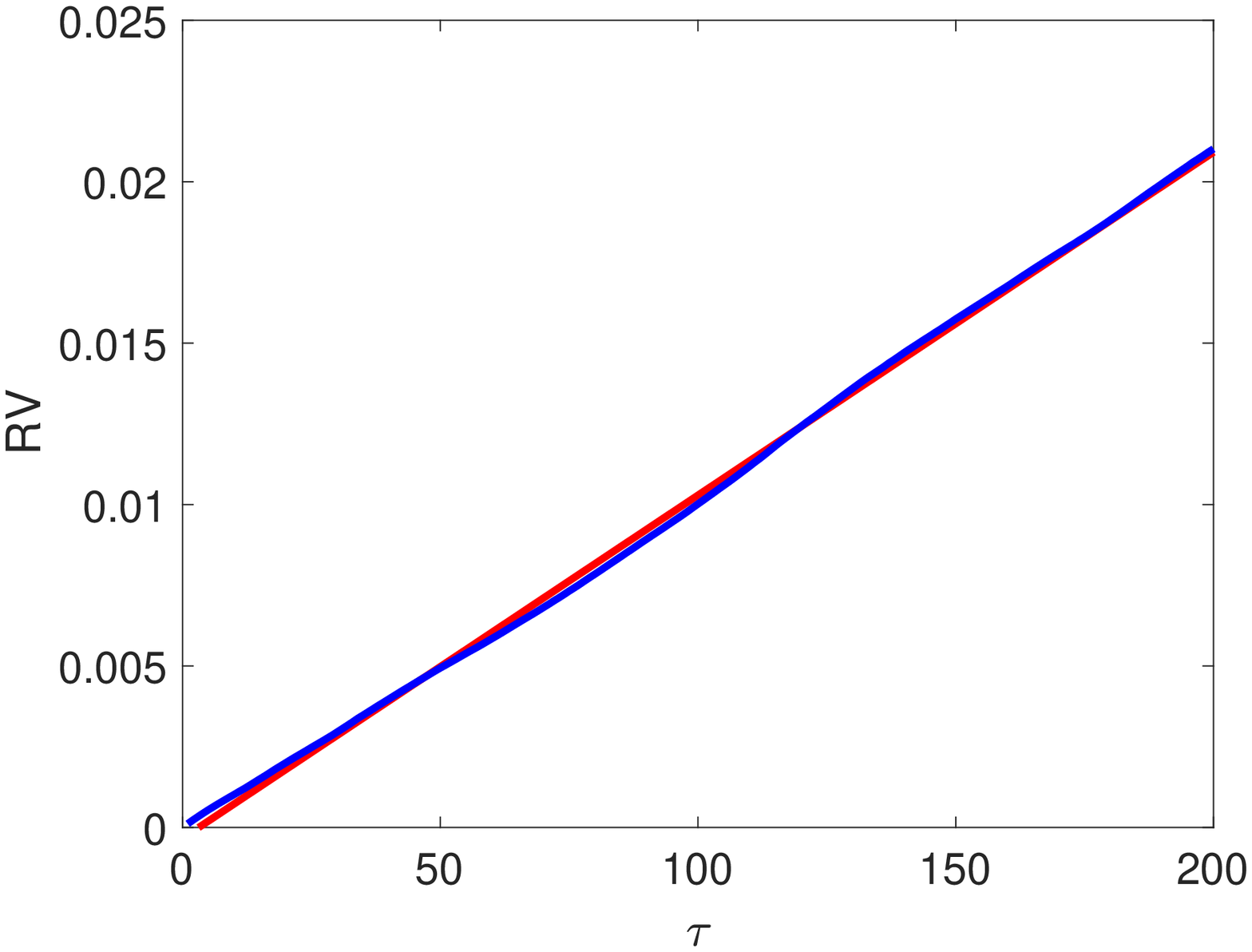}  \\
\end{tabular}

\begin{tabular}{cc}
\includegraphics[width = 0.33 \textwidth]{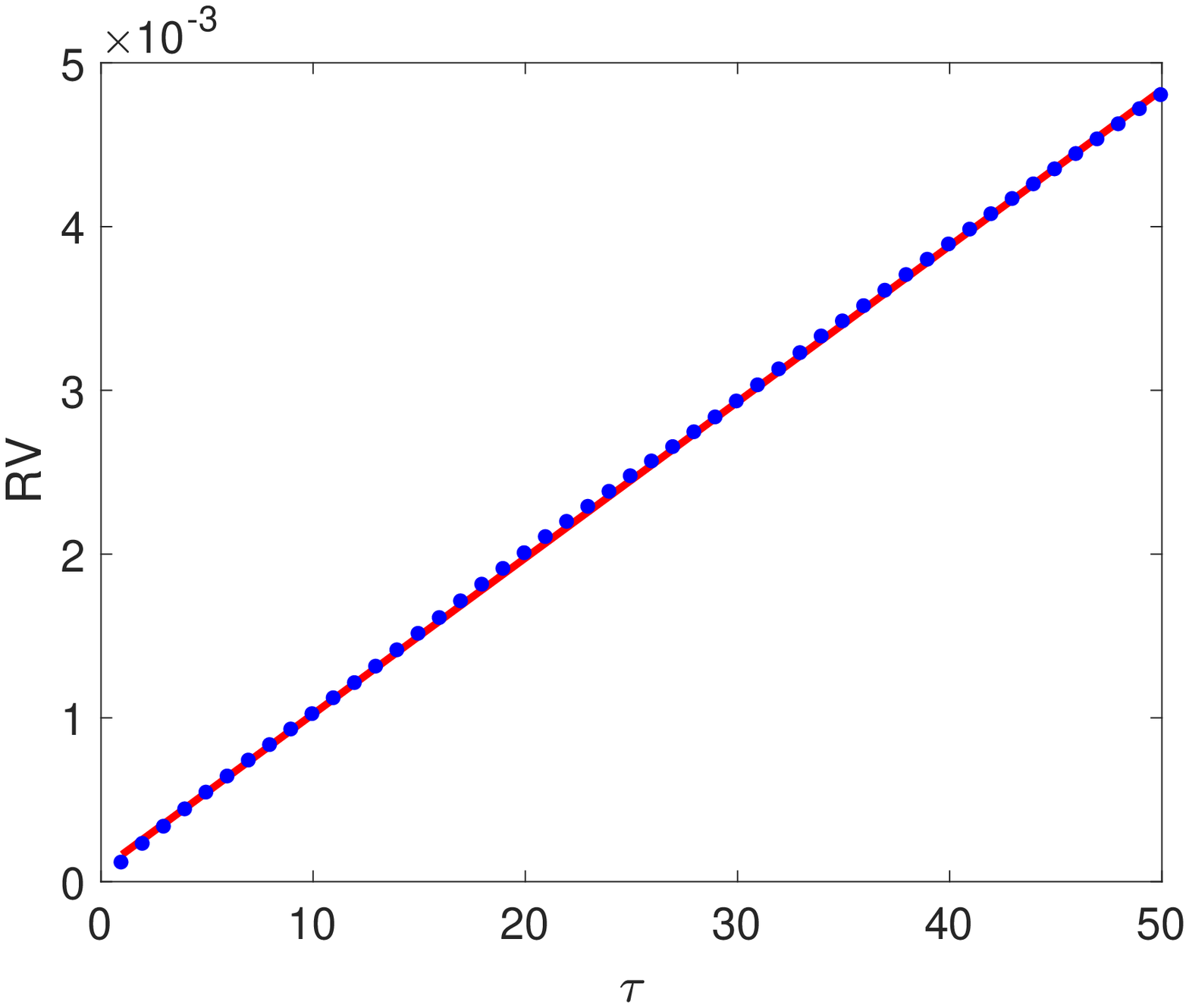}  &
\includegraphics[width = 0.33 \textwidth]{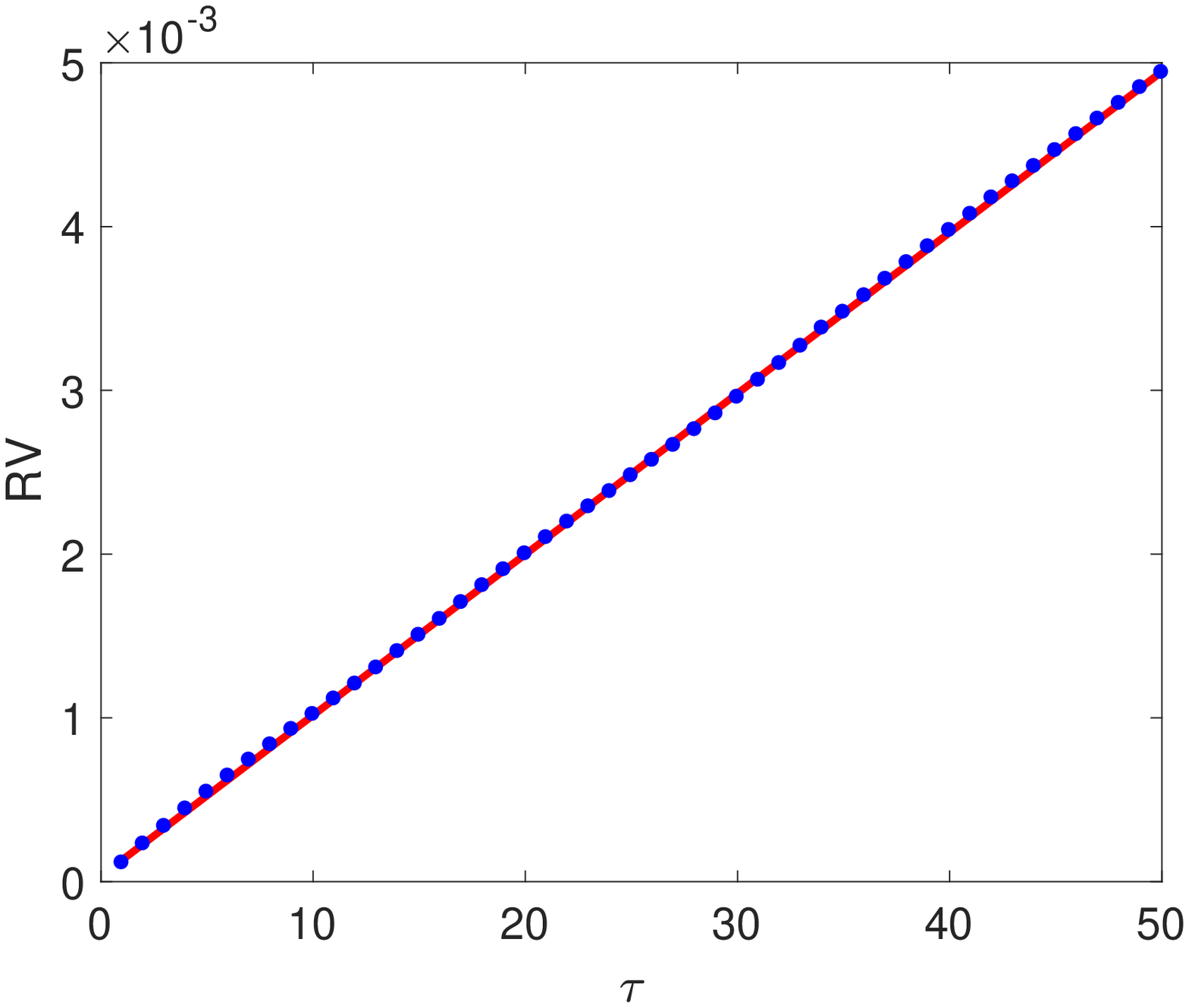} \\
\end{tabular}
\caption{RV of DJIA(left) and S\&P500(right) as a function of the number of days $\tau$ over which the returns are calculated. The best straight line fit for DJIA(left)  is $f(\tau) = 9.578 \times 10^{-5} \tau + 5.283 \times 10^{-5}$ and for S\&P500(right) is $f(\tau) = 1.062 \times 10^{-4} \tau - 3.328 \times 10^{-5}$.}
\label{SPRVWithTau}
\end{figure}

We first concentrate on RV of SR, because its behavior is theoretically independent of the volatility model. Namely, from (\ref{sdrxtapprox}) (and more generally from (\ref{sdrxt}))  RV of SR of $\tau$ days is given by \cite{demeterfi1999more,barndorff2002econometric}
\begin{equation}
V(\tau) = \int_{0}^{\tau}(\mathrm{d}x_t)^2 \approx \int_{0}^{\tau}\sigma_t^2\mathrm{d}t = \int_{0}^{\tau}v_t\mathrm{d}t 
\label{RVdef}
\end{equation}
Integrating (\ref{sdrGaGa}) and (\ref{sdrIGaIGa}), we observe that stochastic terms and  $v_\tau-v_0$ do not contribute on average and so we find 
\begin{equation}
\overline{V(\tau)} = \theta\tau
\label{RVcalc}
\end{equation}
To verify (\ref{RVcalc}) numerically, we use DJIA and S\&P500 data downloaded from Google Finance, as shown in Fig. \ref{GoogleHistoricData}. Fig. \ref{SPRVWithTau} show linear fits of RV as a function of the number of days $\tau$ over which the returns are calculated. These indicate $\theta \approx 9.578 \times 10^{-5}$ for DIJA and $\theta \approx 1.062 \times 10^{-4}$ for S\&P500 -- values very close to values obtained by fitting in Sec. \ref{Numerics} and that in \cite{dragulescu2002probability}.

\section{Derivation of the Distribution Function of Stock Returns \label{DistrSR}}

The main idea in deriving the probability distribution function (PDF) for SR is that if we know the steady state distribution of the stochastic volatility then, according to (\ref{sdrxtapprox}), the distribution of SR is just the PD of the latter, represented by $\sigma_t$, with the normal distribution, represented by $\mathrm{d}W_t^{(1)}$  \cite{ma2014model}. (This can be also cast in terms of conditional probabilities, \cite{praetz1972distribution}). It will be also argued here and in the Appendix that the first term in (\ref{sdrxt}) does not introduce meaningful corrections to the distribution function. 

From (\ref{sdrGaGa}) and (\ref{sdrIGaIGa}) it follows that the steady-state distributions for $v_t$ and $\sigma_t$ are respectively $\alpha \mathrm{Ga(}\alpha v_t;\, \alpha,\, \theta \mathrm{)}$ and $2\sigma_t \cdot \alpha \mathrm{Ga(}\alpha \sigma_t^2;\, \alpha,\, \theta \mathrm{)}$ for the HM  \cite{dragulescu2002probability} and $\mathrm{IGa(}v_t;\, \frac{\alpha }{\theta}+1,\, \alpha \mathrm{)}$ and $2\sigma_t \cdot \mathrm{IGa(}\sigma_t^2;\, \frac{\alpha }{\theta}+1,\, \alpha \mathrm{)}$ for the multiplicative model (MM)\cite{ma2014model}, where $\alpha = \frac{2\gamma \theta}{\kappa^2}$ for both models, with $\alpha > 1$ for the HM. Accordingly, the PDF of the SR for the HM is obtained as 
\begin{equation}
\psi_H(z) = \int_0^{\infty} 2\sigma_t \cdot \alpha \mathrm{Ga(}\alpha \sigma_t^2;\, \alpha,\, \theta \mathrm{)} \cdot \mathrm{N(}z/\sigma_t;\, 0,\, \mathrm{d}t \mathrm{)} \frac{1}{\sigma_t} \mathrm{d}\sigma_t
\label{productH}
\end{equation}
which yields, upon replacing $\mathrm{d}t$ with $\tau$, the following PDF for the $\tau$-day returns:
\begin{equation}
\psi_H(z) = \frac{2^{-\left(\alpha - \frac{1}{2}\right)}}{\sqrt{\pi } \Gamma (\alpha )} \sqrt{\frac{2 \alpha }{\theta  \tau }}   \left(\sqrt{\frac{2 \alpha }{\theta  \tau }} \left| z\right|\right)^{\alpha - \frac{1}{2}} K_{\alpha -\frac{1}{2}}\left(\sqrt{\frac{2 \alpha }{\theta  \tau }} \left| z\right| \right)
\label{pdfH}
\end{equation}
where $K$ is the modified Bessel function of the second kind of order $\alpha-\frac{1}{2}$. For the MM we have
\begin{equation}
\psi_M(z) = \int_0^{\infty} 2\sigma_t \cdot \mathrm{IGa(}\sigma_t^2;\, \frac{\alpha }{\theta}+1,\, \alpha \mathrm{)} \cdot \mathrm{N(}z/\sigma_t;\, 0,\, \mathrm{d}t \mathrm{)} \frac{1}{\sigma_t} \mathrm{d}\sigma_t
\label{productM}
\end{equation}
which yields a Student's distribution  \cite{praetz1972distribution,ma2014model}
\begin{equation}
\psi_M(z) = \frac{\Gamma \left(\frac{\alpha }{\theta }+\frac{3}{2}\right)}{\sqrt{\pi } \Gamma \left(\frac{\alpha }{\theta }+1\right)} \frac{1}{\sqrt{2 \alpha  \tau }} \left(\frac{z^2}{2 \alpha  \tau }+1\right)^{-\left(\frac{\alpha }{\theta }+\frac{3}{2}\right)}
\label{pdfM}
\end{equation}

The two immediate observations follow from (\ref{pdfH}) and (\ref{pdfM}): first, that both PDFs scale with $\tau^{1/2}$ and second, that the variance of the distribution ($\tau$-days RV) is given by 
\begin{equation}
\int_{-\infty}^{\infty} z^2 \psi_H (z) \mathrm{d}z = \int_{-\infty}^{\infty} z^2 \psi_M (z) \mathrm{d}z = \theta \tau
\label{RVtau}
\end{equation}
in agreement with (\ref{RVcalc}).

The first term in the r.h.s. of (\ref{sdrxt}) can be taken into account using joint probability (JP) argument (see Appendix), which results in 
\begin{equation}
\phi_H(z) = \frac{2^{-\left(\alpha - \frac{1}{2}\right)}}{\sqrt{\pi } \Gamma (\alpha )} 
\left(\frac{2 \alpha }{\theta  \tau }\right)^{\alpha } \left(\frac{2 \alpha }{\theta  \tau }+\frac{1}{4}\right)^{-\left(\alpha -\frac{1}{2}\right)}
\left(\sqrt{\frac{2 \alpha }{\theta  \tau }+\frac{1}{4}} \left| z\right| \right)^{\alpha -\frac{1}{2}}
K_{\alpha -\frac{1}{2}}\left(\sqrt{\frac{2 \alpha }{\theta  \tau }+\frac{1}{4}} \left| z\right|\right) e^{-\frac{z}{2}}
\label{pdfJH}
\end{equation}
and
\begin{equation}
\phi_M(z) = \frac{2^{-2 \left(\frac{\alpha }{\theta }+1\right)}}{\sqrt{\pi } \Gamma \left(\frac{\alpha }{\theta }+1\right)}
\left(\sqrt{2 \alpha  \tau }\right)^{\frac{\alpha }{\theta }+\frac{1}{2}}
\left(\sqrt{\frac{z^2}{2 \alpha  \tau }+1}\right)^{-\left(\frac{\alpha }{\theta }+\frac{3}{2}\right)}
K_{\frac{\alpha }{\theta }+\frac{3}{2}}\left(\frac{\sqrt{2 \alpha  \tau }}{2} \sqrt{\frac{z^2}{2 \alpha  \tau }+1}\right)
e^{-\frac{z}{2}}
\label{pdfJM}
\end{equation}
for Heston and multiplicative models respectively.

In \cite{dragulescu2002probability}, the full time-dependent Fokker-Planck equations was solved. Upon inspection, one finds that (\ref{pdfJH}) is a slightly more accurate expression (49) for the ``short-time'' limit there. Furthermore, given that the ``short-time'' limit in \cite{dragulescu2002probability} is defined as $\frac{2 \alpha }{\theta  \tau } \gg \frac{1 }{4 }$, that is $\theta \tau \ll 8\alpha$, we find that in this limit (\ref{pdfJH}) reduces to (\ref{pdfH}) with the exception of $e^{-\frac{z}{2}}$ in (\ref{pdfJH}). However, the relevant scale for $z$ is defined by the width of the distribution, $(\theta \tau)^{1/2}$. Consequently, when $\theta \tau \ll 1$ the exponent in (\ref{pdfJH}) can be neglected and in this limit (\ref{pdfJH}) completely reduces to (\ref{pdfH}). We also point out that the short-time limit in \cite{dragulescu2002probability} is $\rho$-independent and so the empirical observation by the authors that over relevant periods of returns $\rho$ can be set to zero is entirely consistent with that one is only interested in the ``short-time'' limit, which can extend to very long periods over which the returns are calculated.

It should be emphasized that a small RV, $\theta \tau \ll 1$, emerges as the universal condition of applicability of PD for the mean-reverting models. For instance, in the MM, one finds the same exponent for JP in (\ref{pdfJM}) as in (\ref{pdfJH}). (The reduction of (\ref{pdfJM}) to (\ref{pdfM}) is less obvious and will be discussed in the Appendix.)  Given that $\theta  \approx 10^{-4}$ (see next Section), we find that PD works very well over very long periods of returns -- measured in years. 

Clearly, (\ref{pdfJH}) and (\ref{pdfJM}) move the SR distribution slightly off-center and introduce a slight skew. However, in the long-time limit one still expects the distribution to approach normal since it reflects the sum of a large number of identical independently distributed (i.i.d.) variables, as follows from (\ref{sdrxtapprox}). Indeed, in \cite{dragulescu2002probability} it is shown that the SR distribution tends to normal in the long-time limit.

\section{Numerical Results\label{Numerics}}

\begin{figure}[!htbp]
\centering
\begin{tabular}{cc}
\includegraphics[width = \myFigureWidth \textwidth]{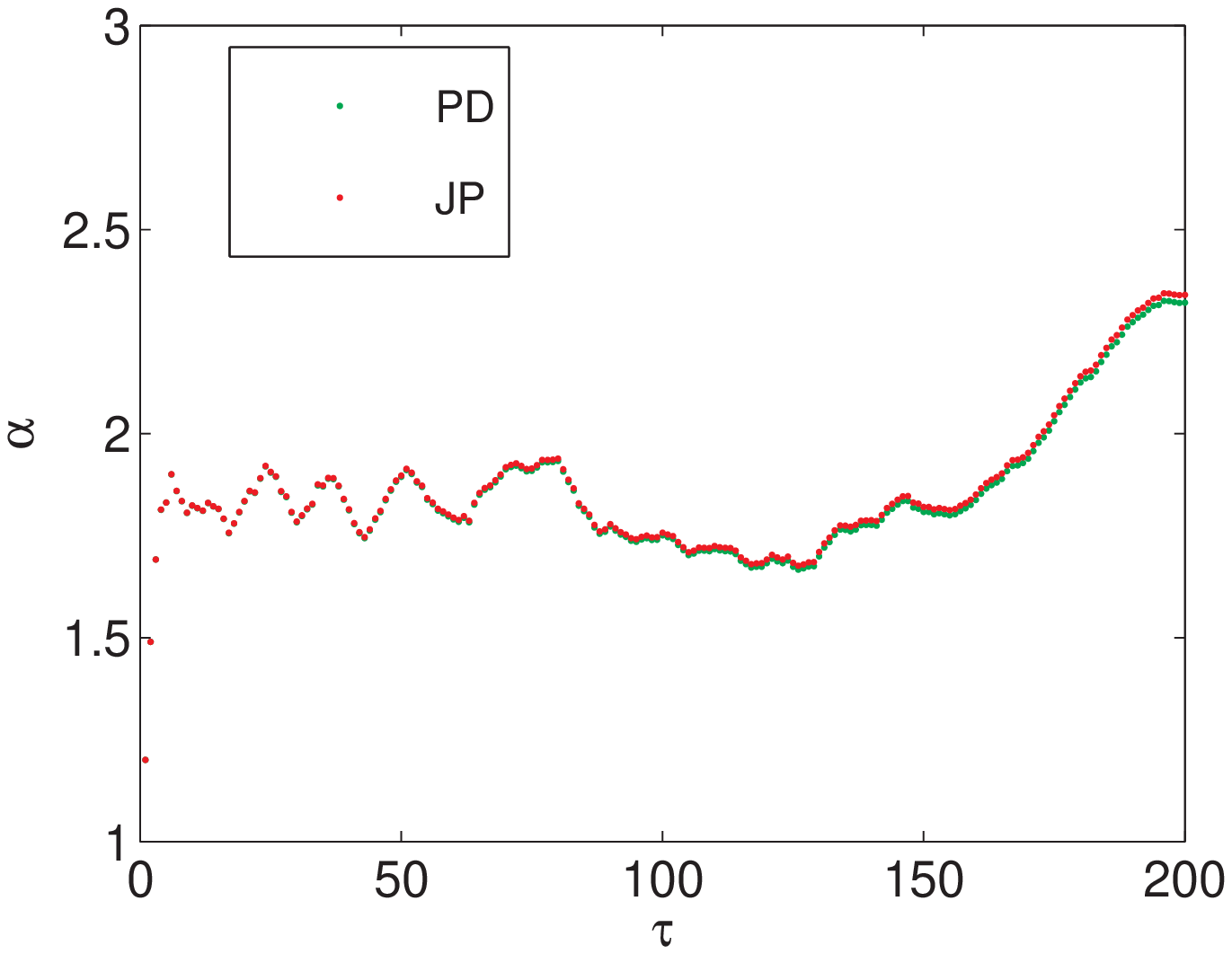} &
\includegraphics[width = 0.32 \textwidth]{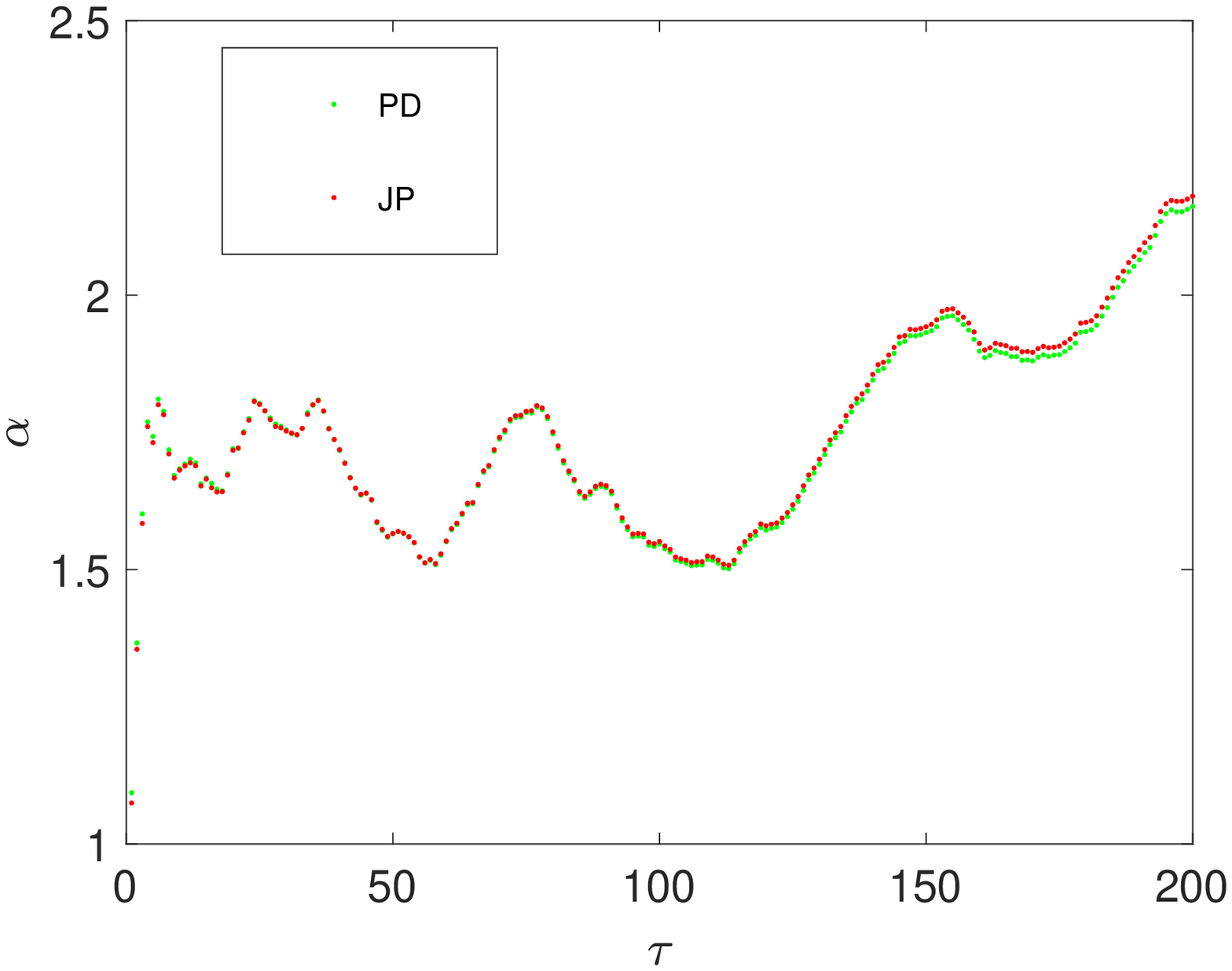} \\
\end{tabular}

\begin{tabular}{cc}
\includegraphics[width = \myFigureWidth \textwidth]{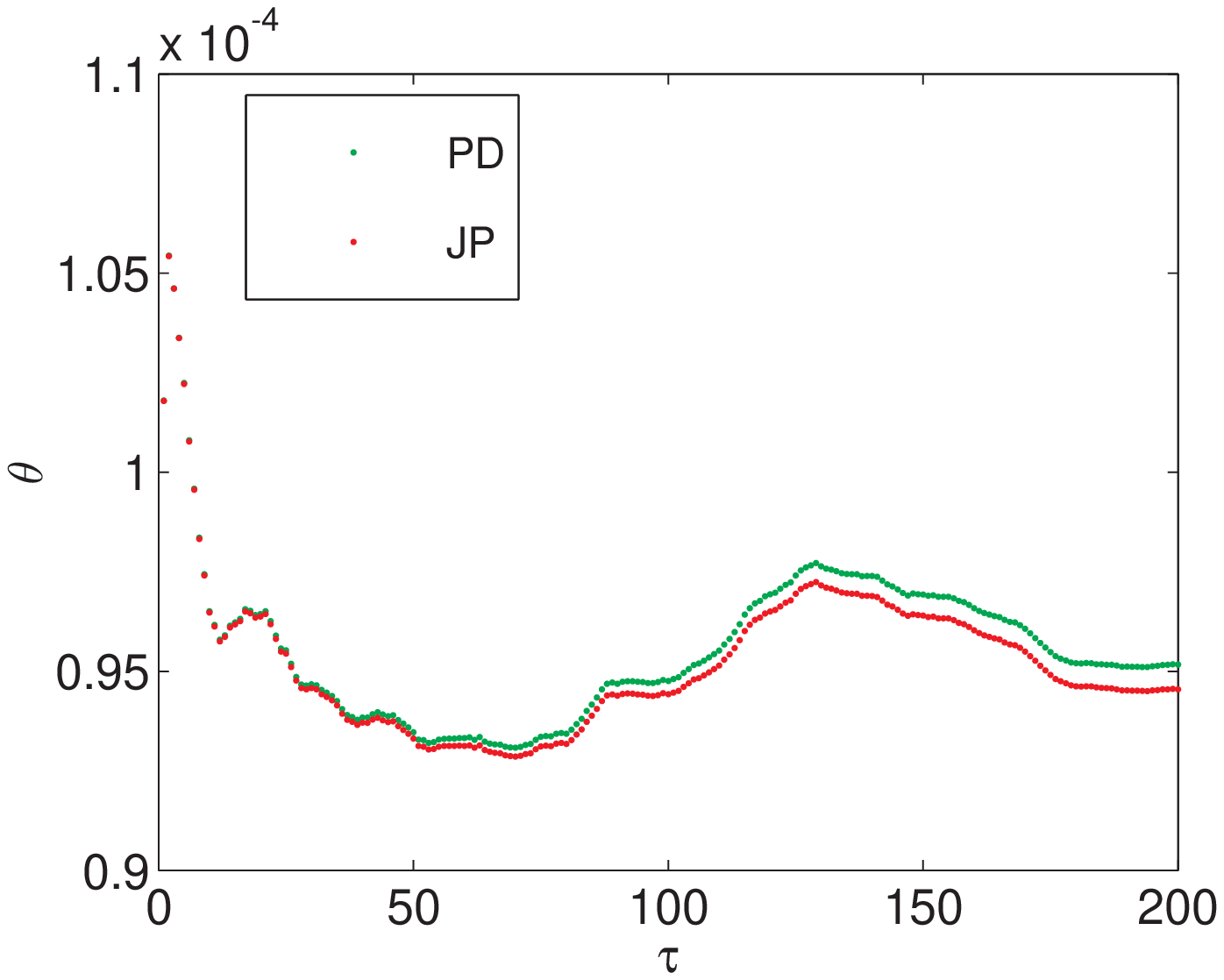} &
\includegraphics[width = 0.32 \textwidth]{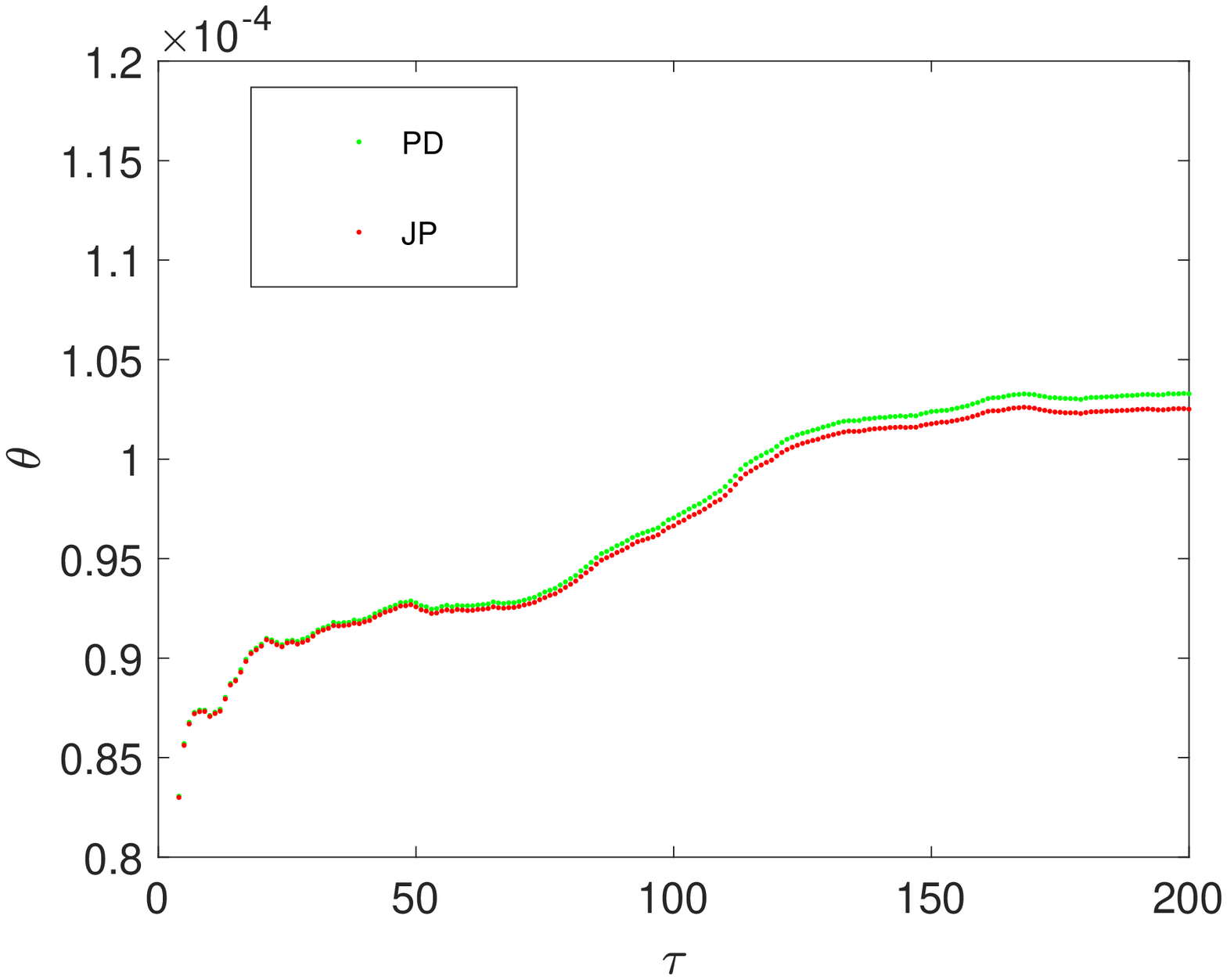} \\
\end{tabular}
\caption{Heston model: Parameters of (\ref{pdfH}) and (\ref{pdfJH}) for DJIA(left) and $\text{S\&P500}$(right)}
\label{VYSPGoogleAllalphaList}
\end{figure}

\begin{figure}[!htbp]
\centering
\begin{tabular}{cc}
\includegraphics[width = \myFigureWidth \textwidth]{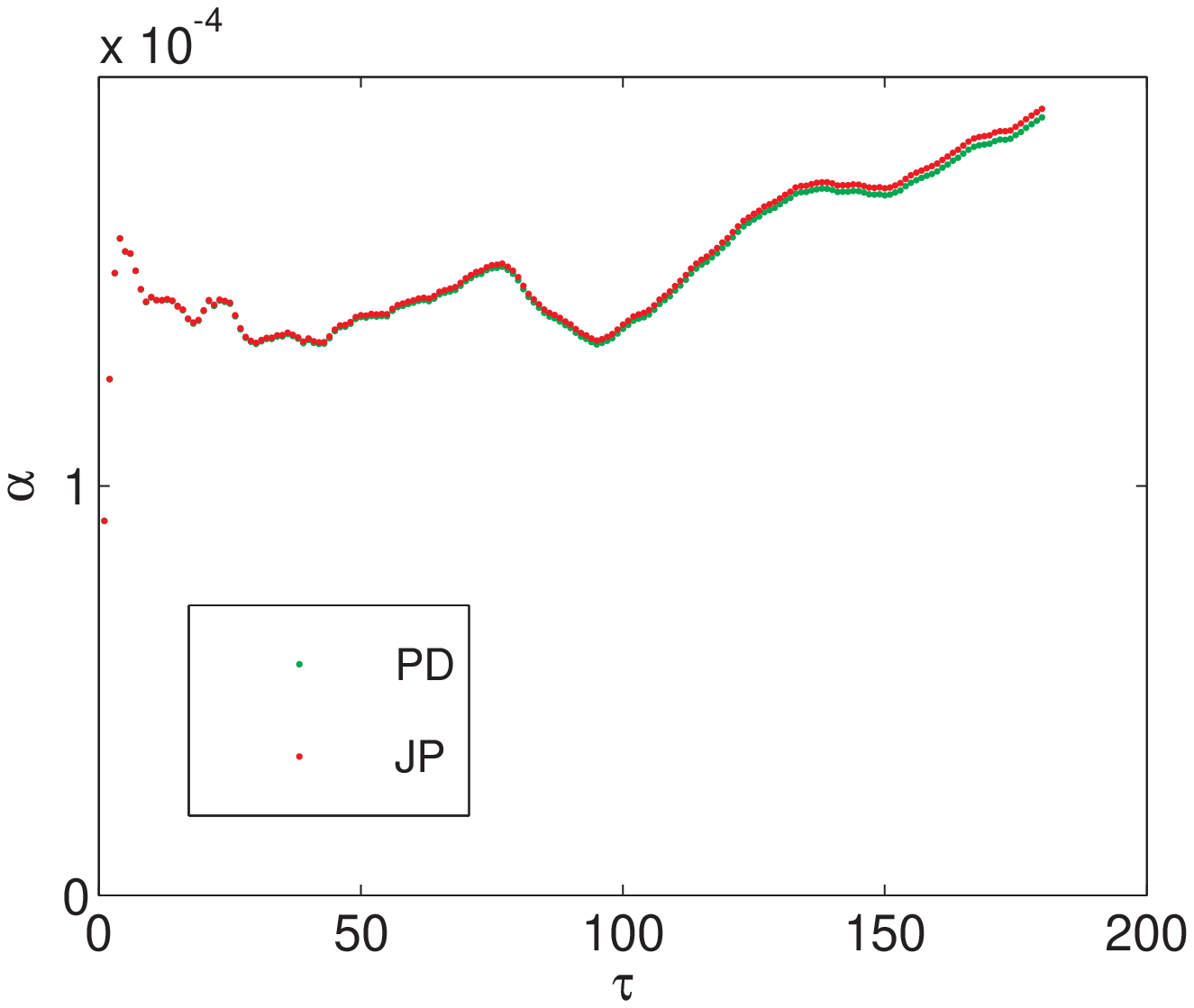} &
\includegraphics[width = 0.33 \textwidth]{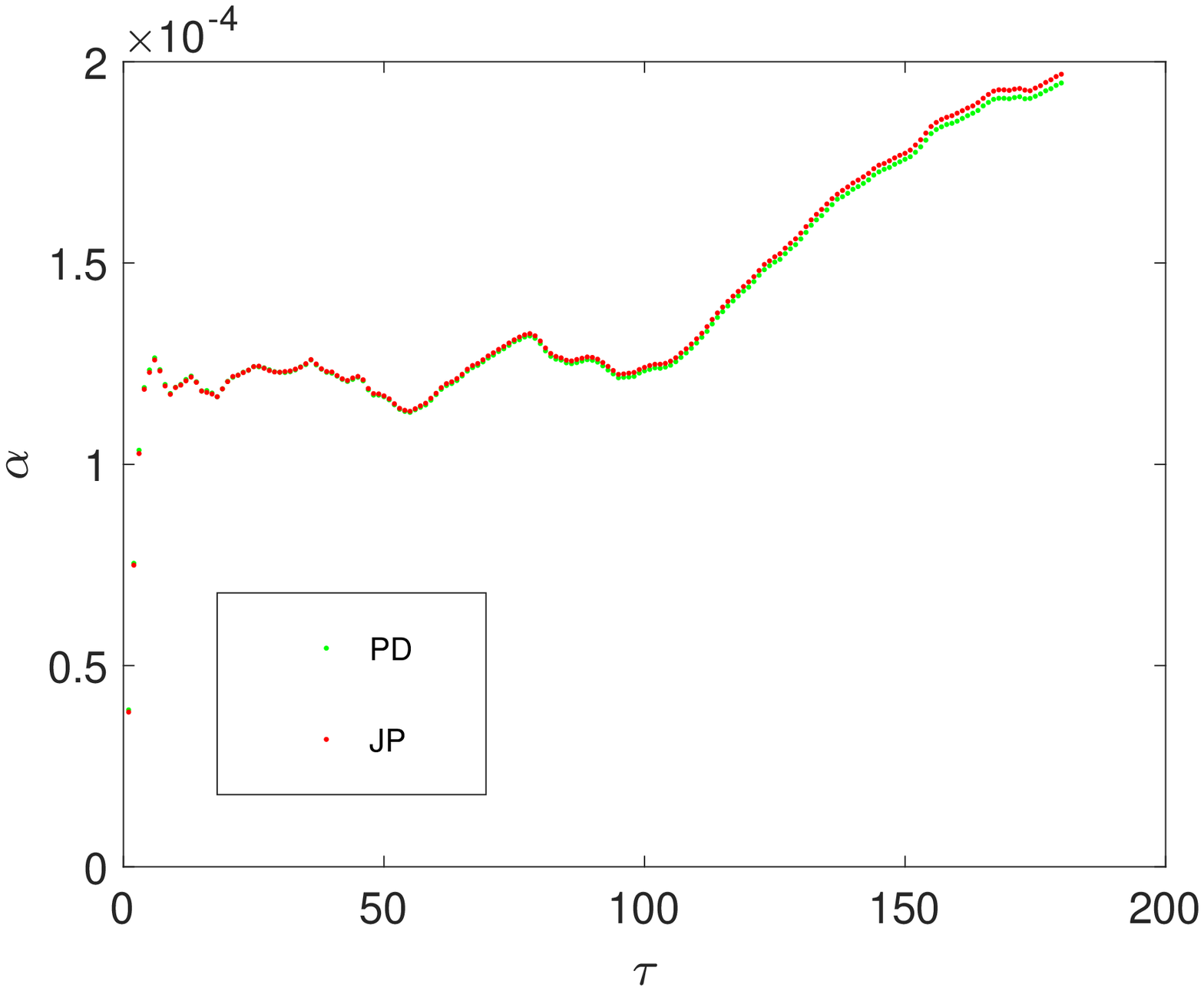} \\
\end{tabular}
\begin{tabular}{cc}
\includegraphics[width = \myFigureWidth \textwidth]{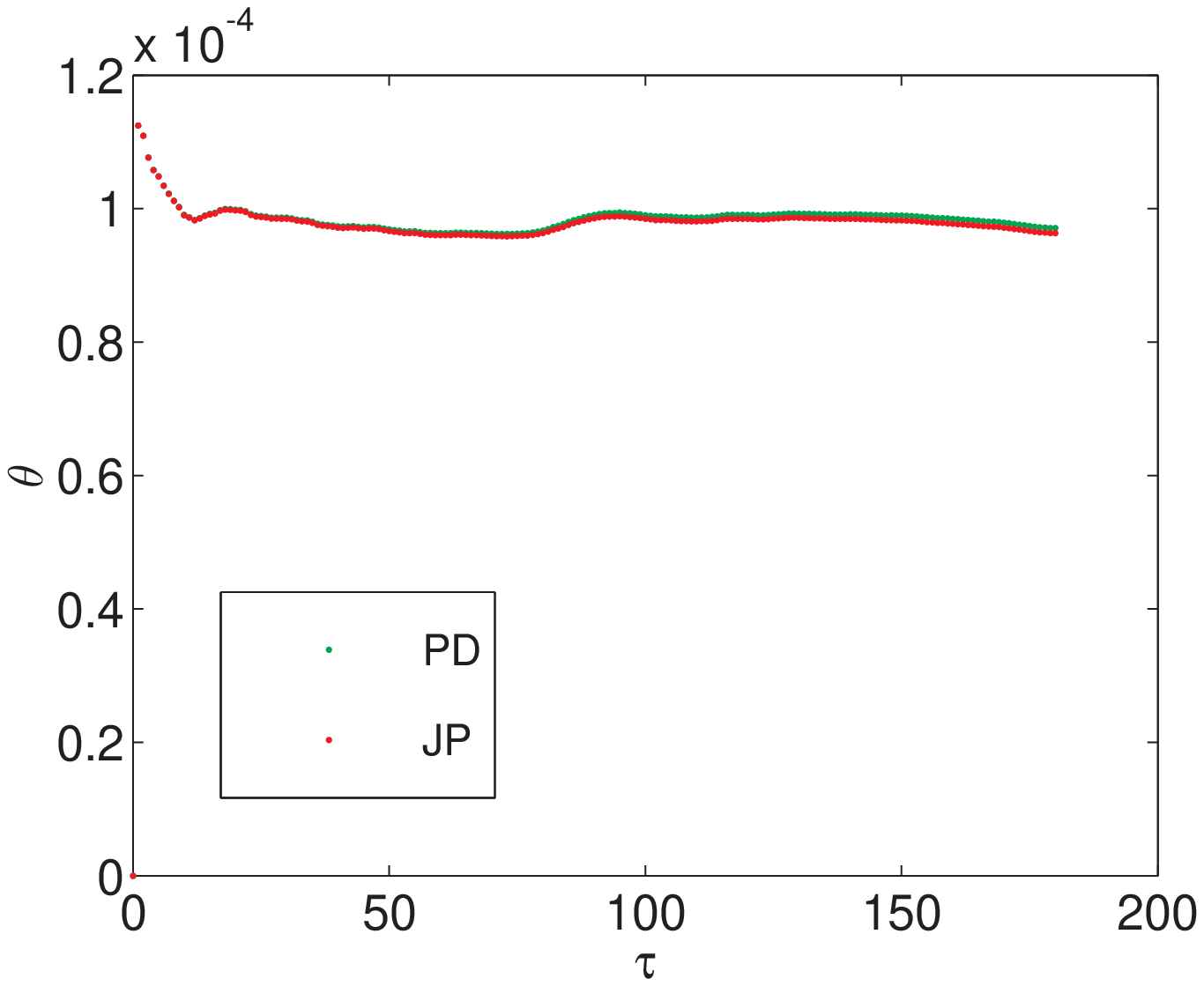}  &
\includegraphics[width = 0.33 \textwidth]{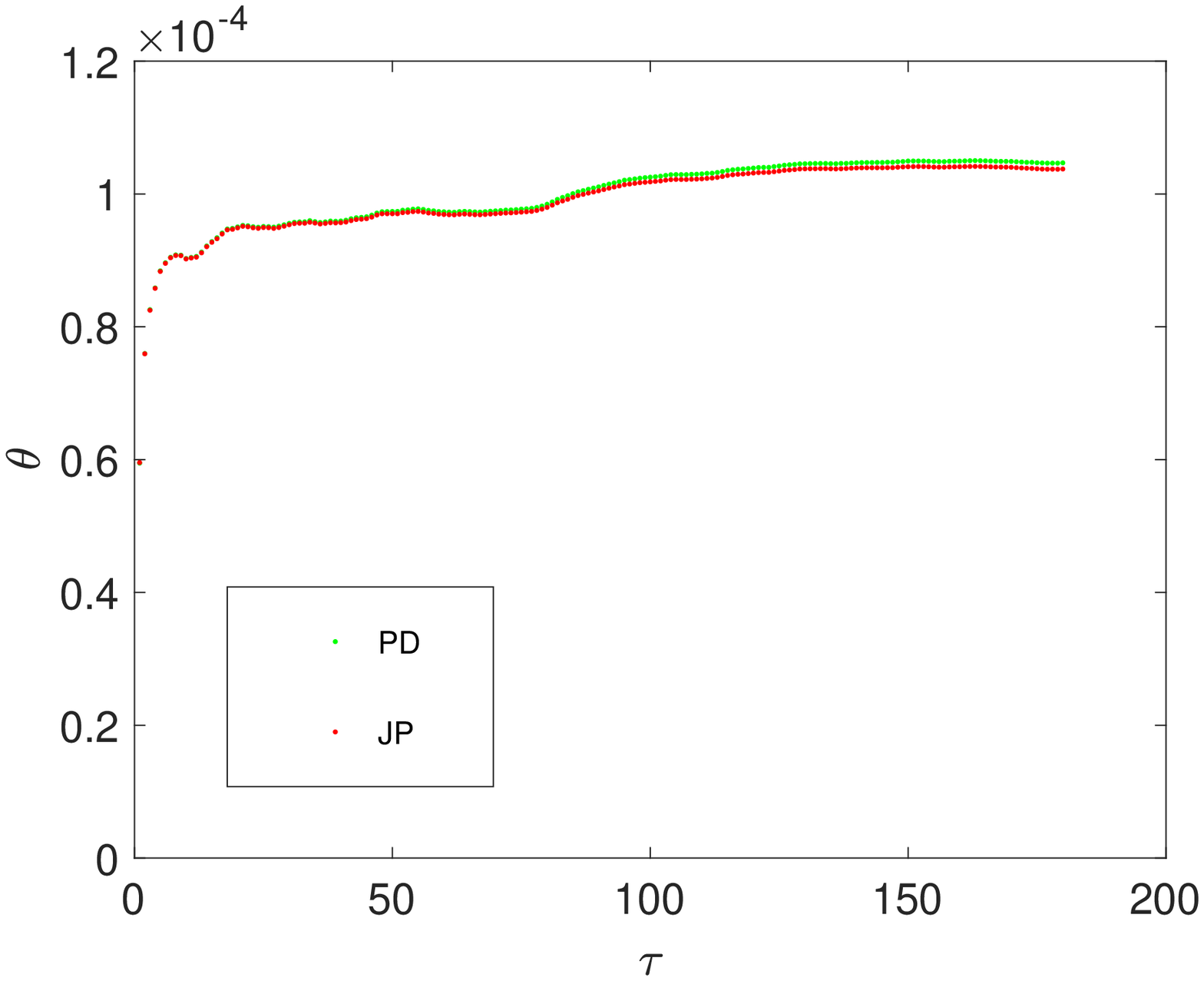} \\
\end{tabular}
\caption{Multiplicative model: Parameters of (\ref{pdfM}) and (\ref{pdfJM}) for DJIA(left) and $\text{S\&P500}$(right)}
\label{UYSPIAGoogleAllalphaList}
\end{figure}

We fit SR distribution with distributions (\ref{pdfH}), (\ref{pdfM}), (\ref{pdfJH}) and (\ref{pdfJM}) using maximum likelihood estimation (MLE) as a function of $\tau$. Results for the HM, (\ref{pdfH}) and (\ref{pdfJH}), are plotted in Fig. \ref{VYSPGoogleAllalphaList}. Results for the Multiplicative  model, (\ref{pdfM}) and (\ref{pdfJM}), are plotted in Fig. \ref{UYSPIAGoogleAllalphaList}. As expected, due to the results of preceding Section, for either model there is very little difference between PD and JP fits. 

It should be pointed out, that since $\alpha = \frac{2\gamma \theta}{\kappa^2}$, a separate consideration is required to find $\gamma$ and $\kappa^2$ separately. Leverage   \cite{bouchaud2001leverage,perello2002correlated,perello2004multiple} can be used to determine $\gamma$ (in addition to $\rho$). We will also see in the next Section that relaxation of the moments of the SR distribution can be used for this purpose as well.

\begin{figure}[!htbp]
\centering
\begin{tabular}{cc}
\includegraphics[height = 0.28 \textwidth]{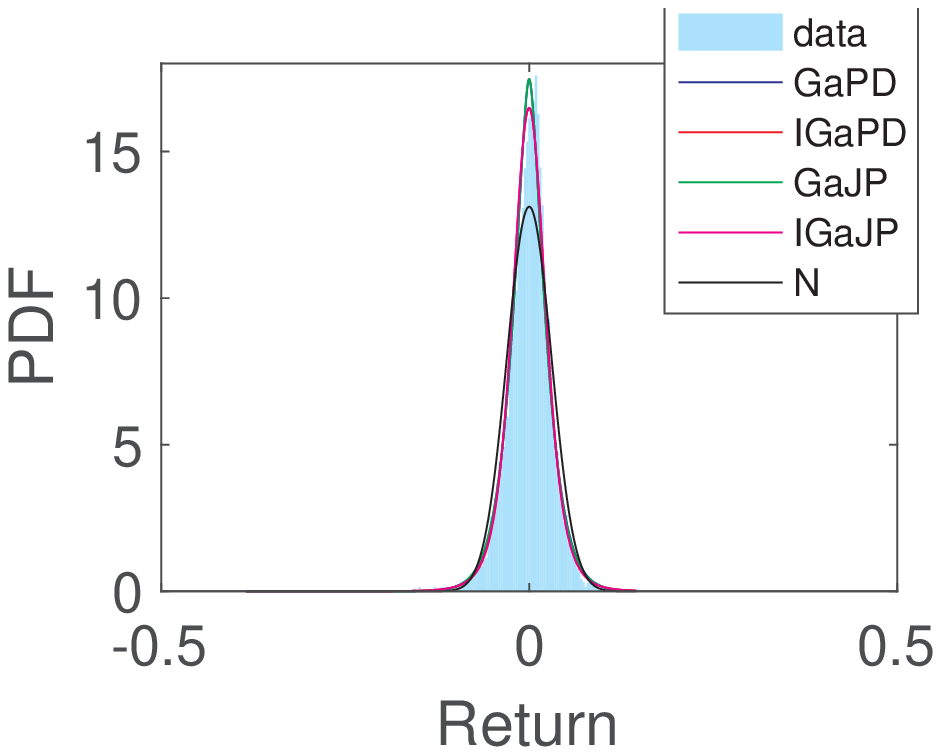}
\includegraphics[height = 0.28 \textwidth]{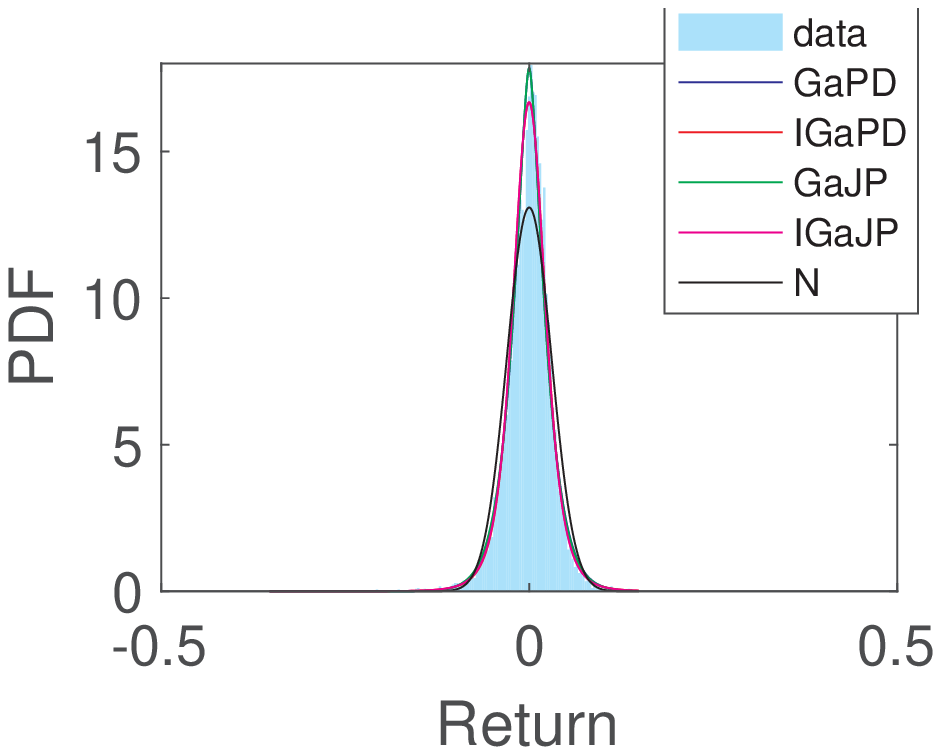}
\end{tabular}
\begin{tabular}{cc}
\includegraphics[height = 0.28\textwidth]{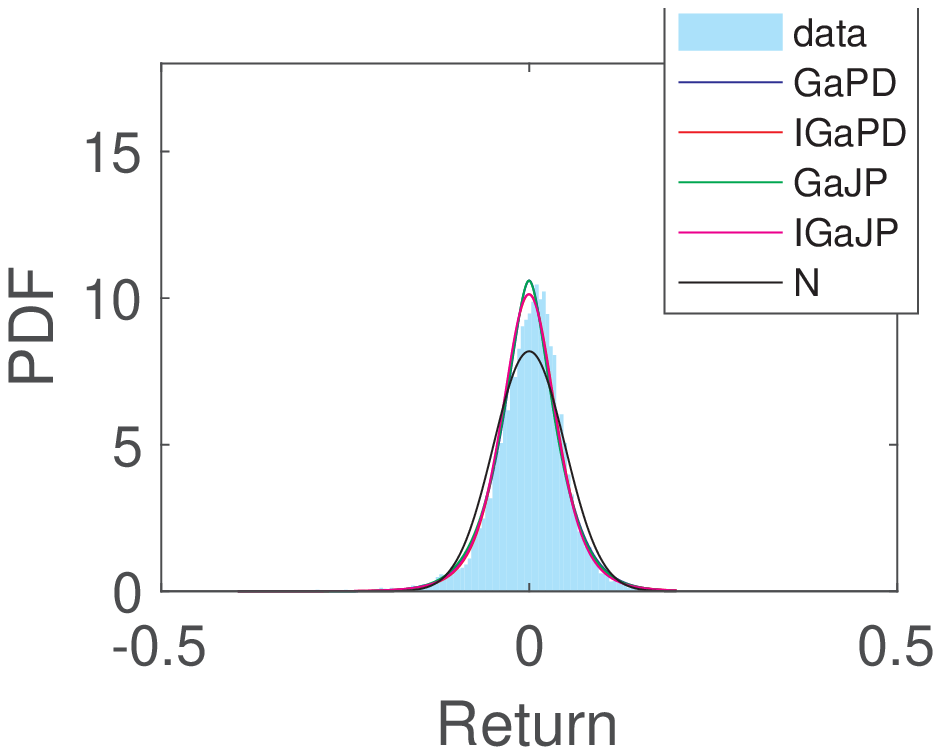}
\includegraphics[height = 0.28\textwidth]{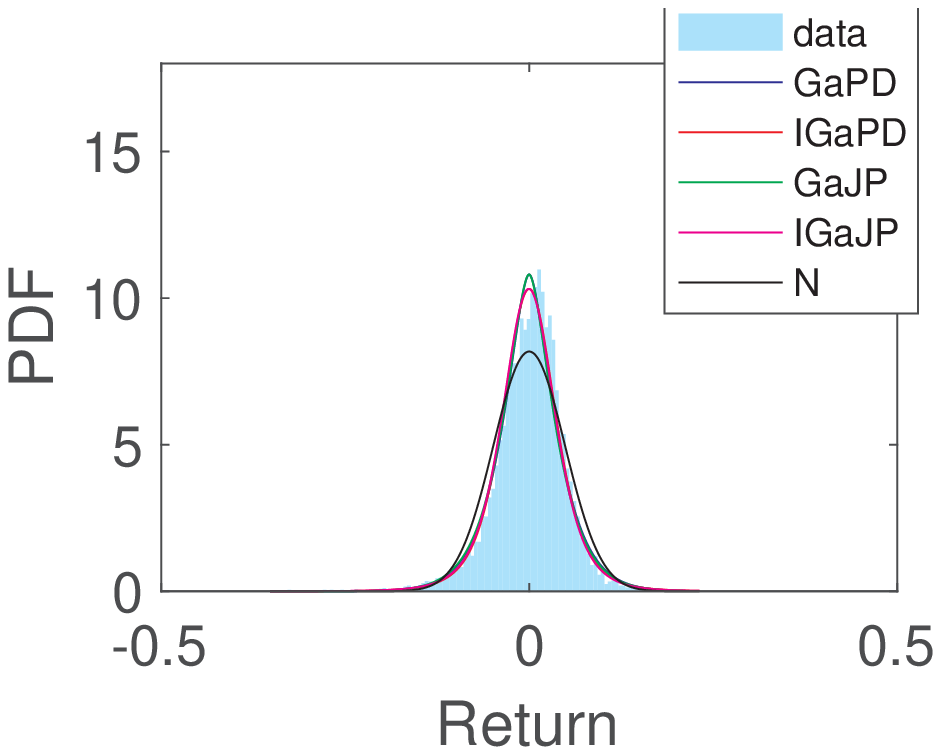}
\end{tabular}
\begin{tabular}{cc}
\includegraphics[height = 0.28 \textwidth]{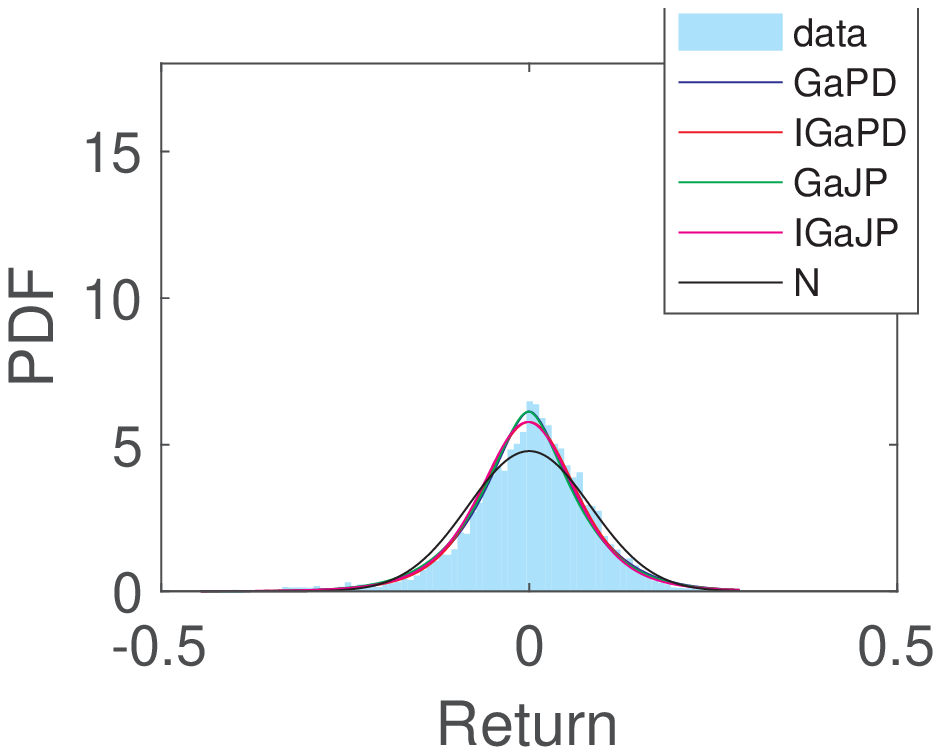}
\includegraphics[height = 0.28\textwidth]{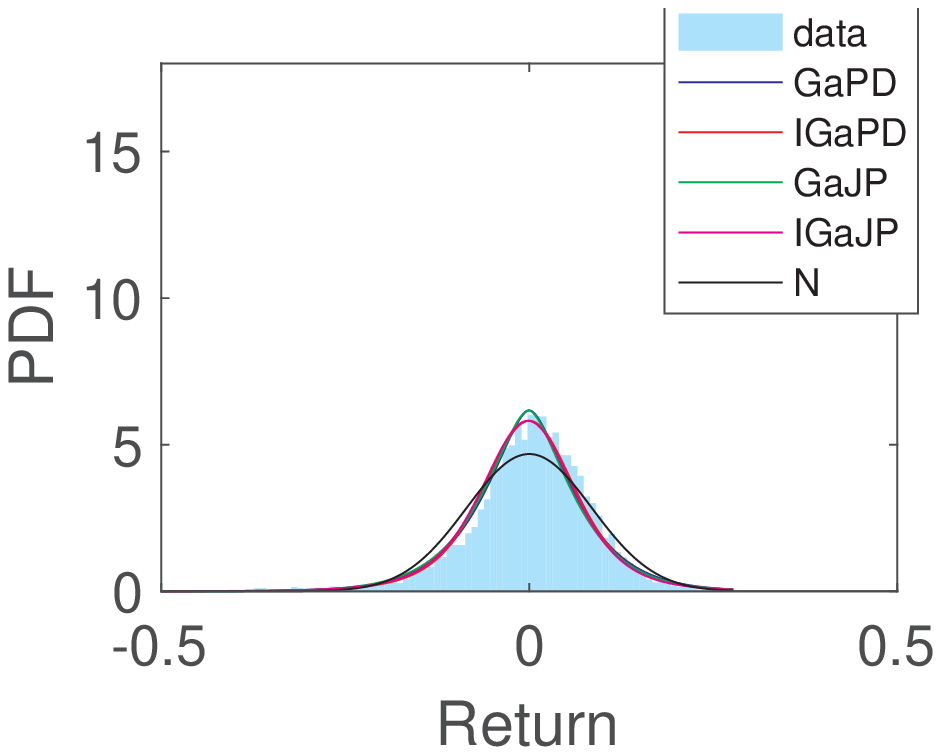}
\end{tabular}
\begin{tabular}{cc}
\includegraphics[height = 0.28 \textwidth]{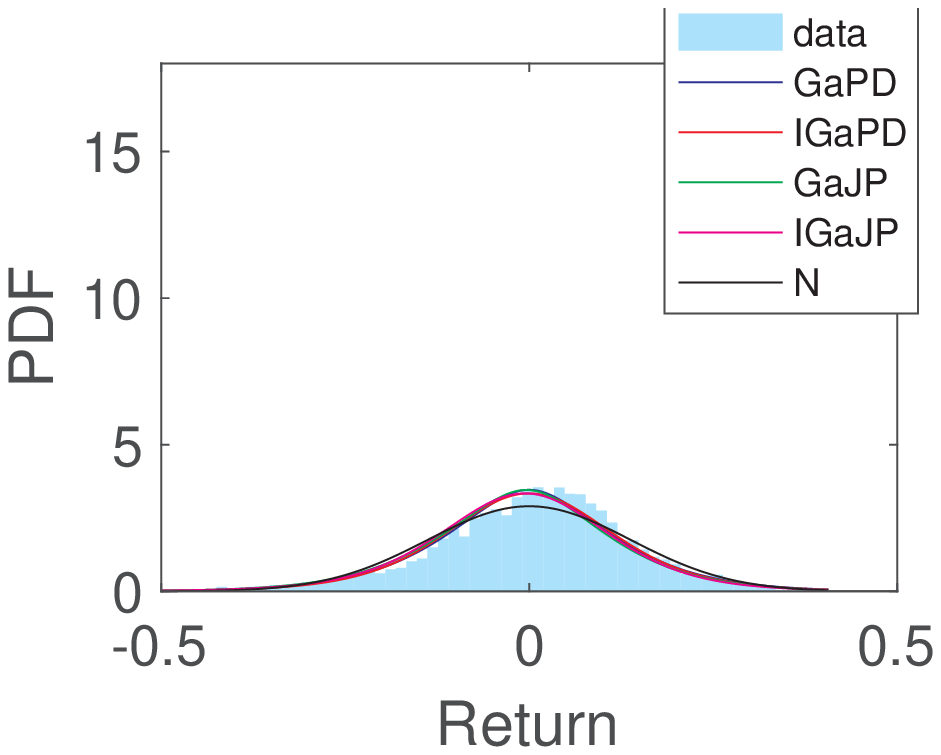}
\includegraphics[height = 0.28 \textwidth]{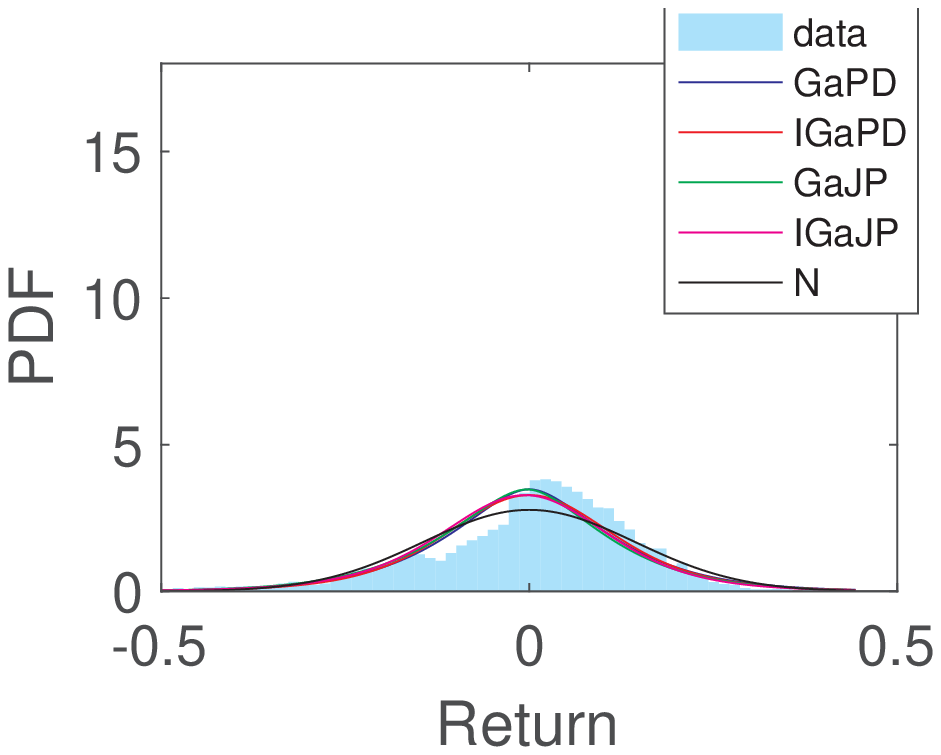}

\end{tabular}
\caption{From top to bottom, PDF of SR for $\tau=10$, 25, 75 and 200 respectively, for DJIA (left) and S\&P500 (right).}
\label{Probability Distribution of Stock Returns}
\end{figure}

Fig. \ref{Probability Distribution of Stock Returns} demonstrates scaling discussed in the preceding section. SR distributions and their fits (\ref{pdfH}), (\ref{pdfJH}), (\ref{pdfM}) and (\ref{pdfJM}), as well as the normal (N) distribution fit, are shown for $\tau=\text {10, 25, 75, and 200 days}$. Visually, there is very little difference between the four, while all of them fit much better than N.

\begin{figure}[!htbp]
\centering
\begin{tabular}{cc}
\includegraphics[width = 0.4 \textwidth]{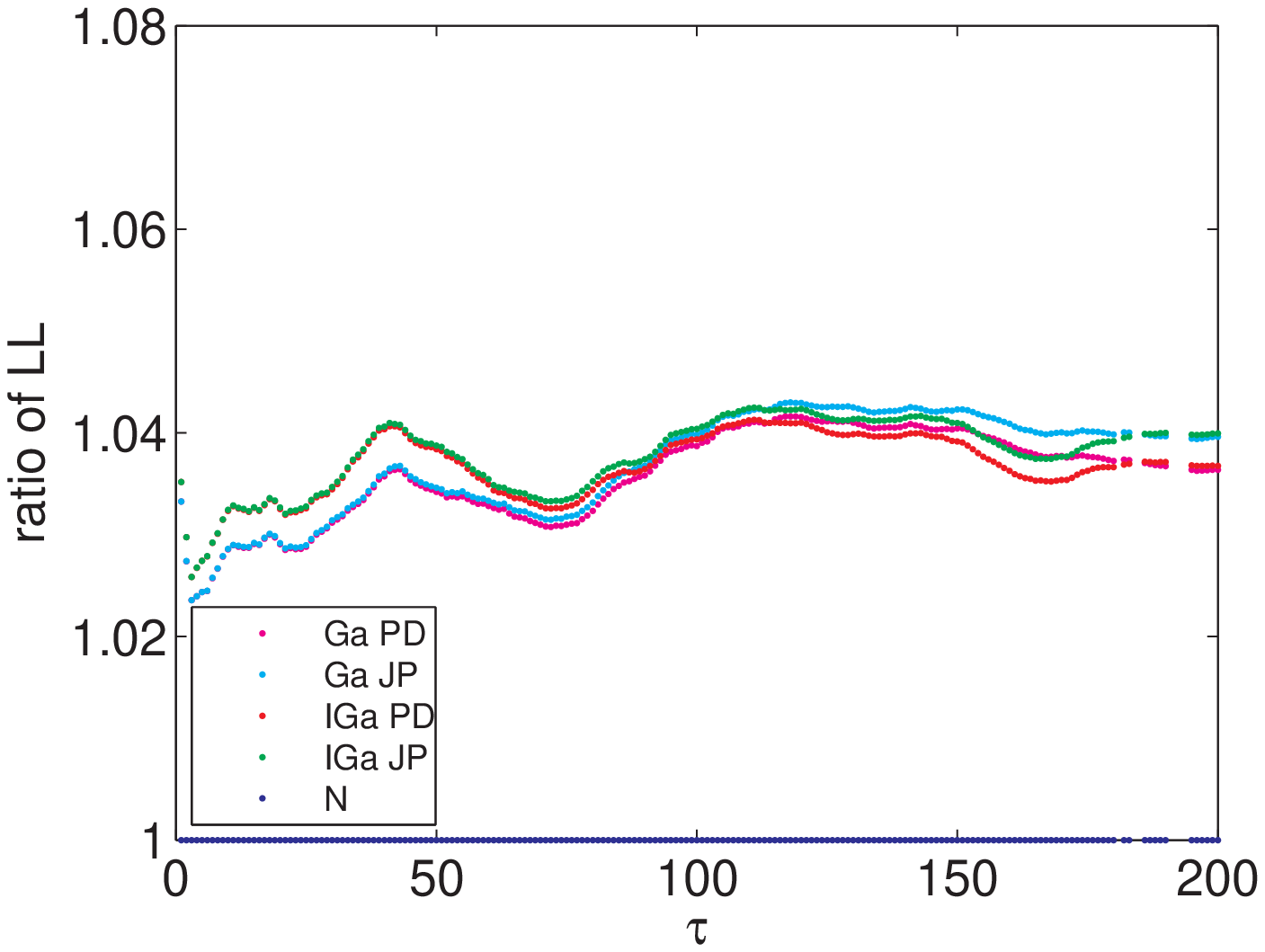}
\includegraphics[width = 0.4 \textwidth]{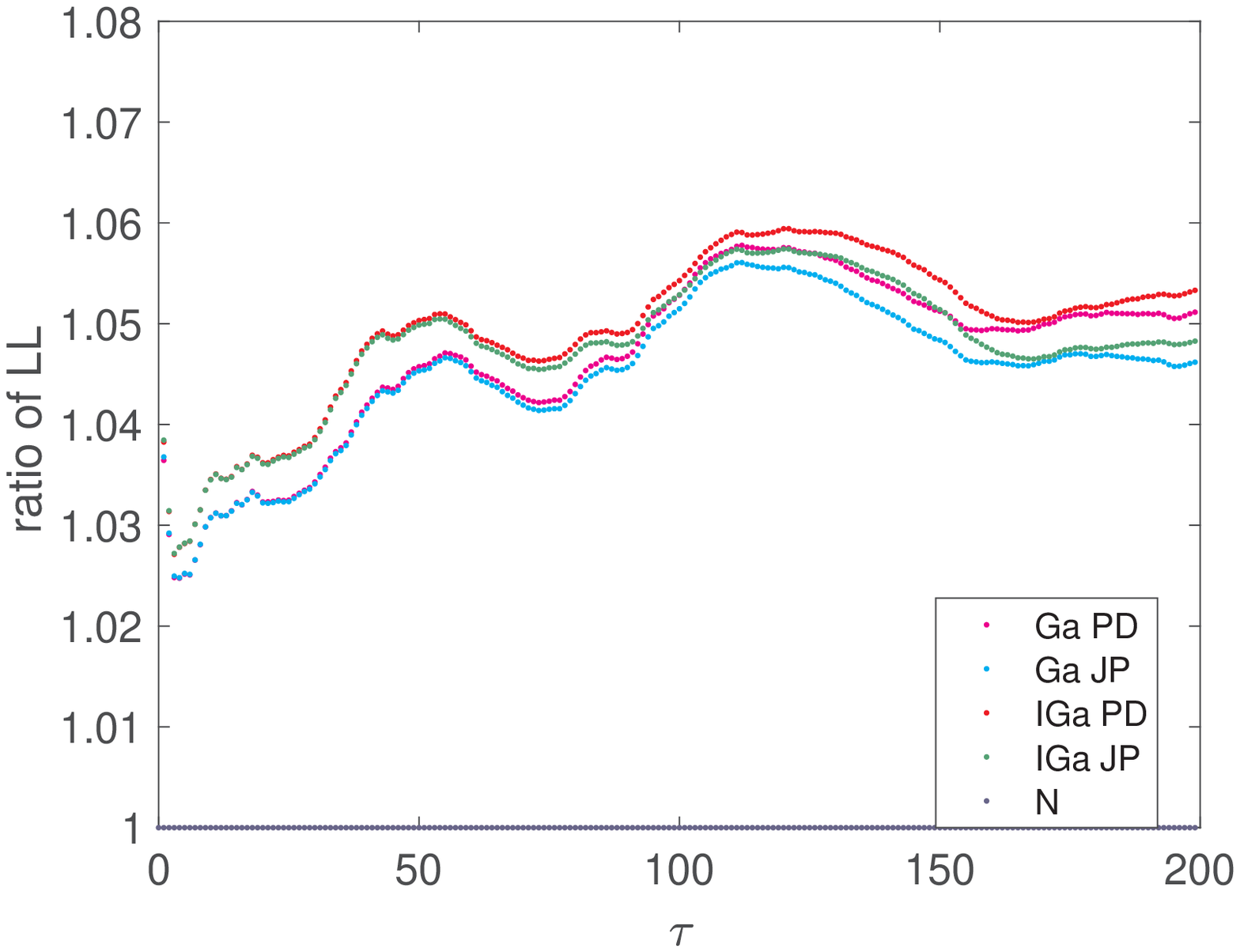}
\end{tabular}

\caption{LL ratio to N for DJIA (left) and S\&P500 (right); ``Ga'' -- Heston model, ``IGa'' -- multiplicative model.}
\label{LLratio}
\end{figure}

\begin{figure}[!htbp]
\centering
\begin{tabular}{cc}
\includegraphics[width = 0.4 \textwidth]{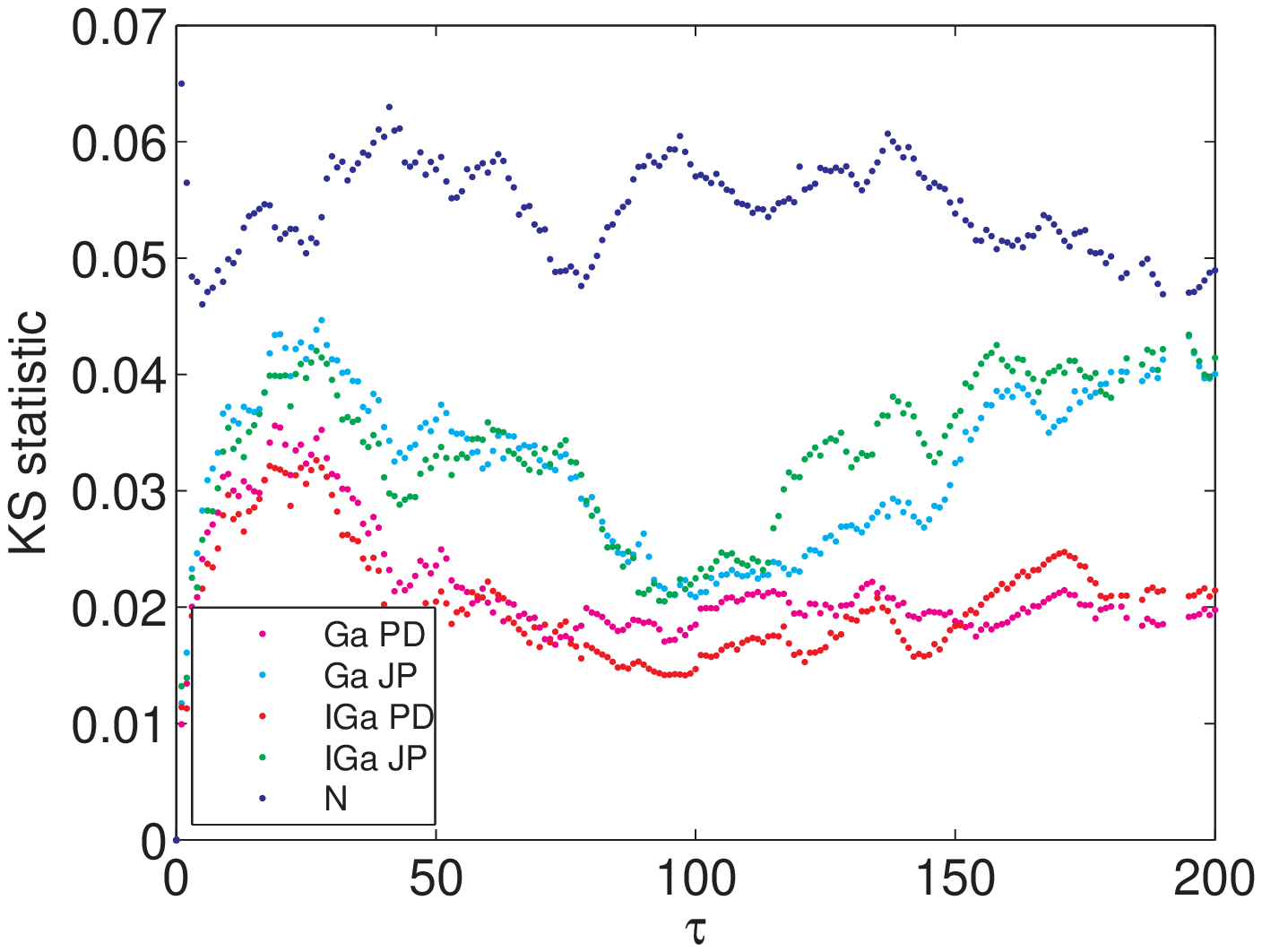}
\includegraphics[width = 0.4 \textwidth]{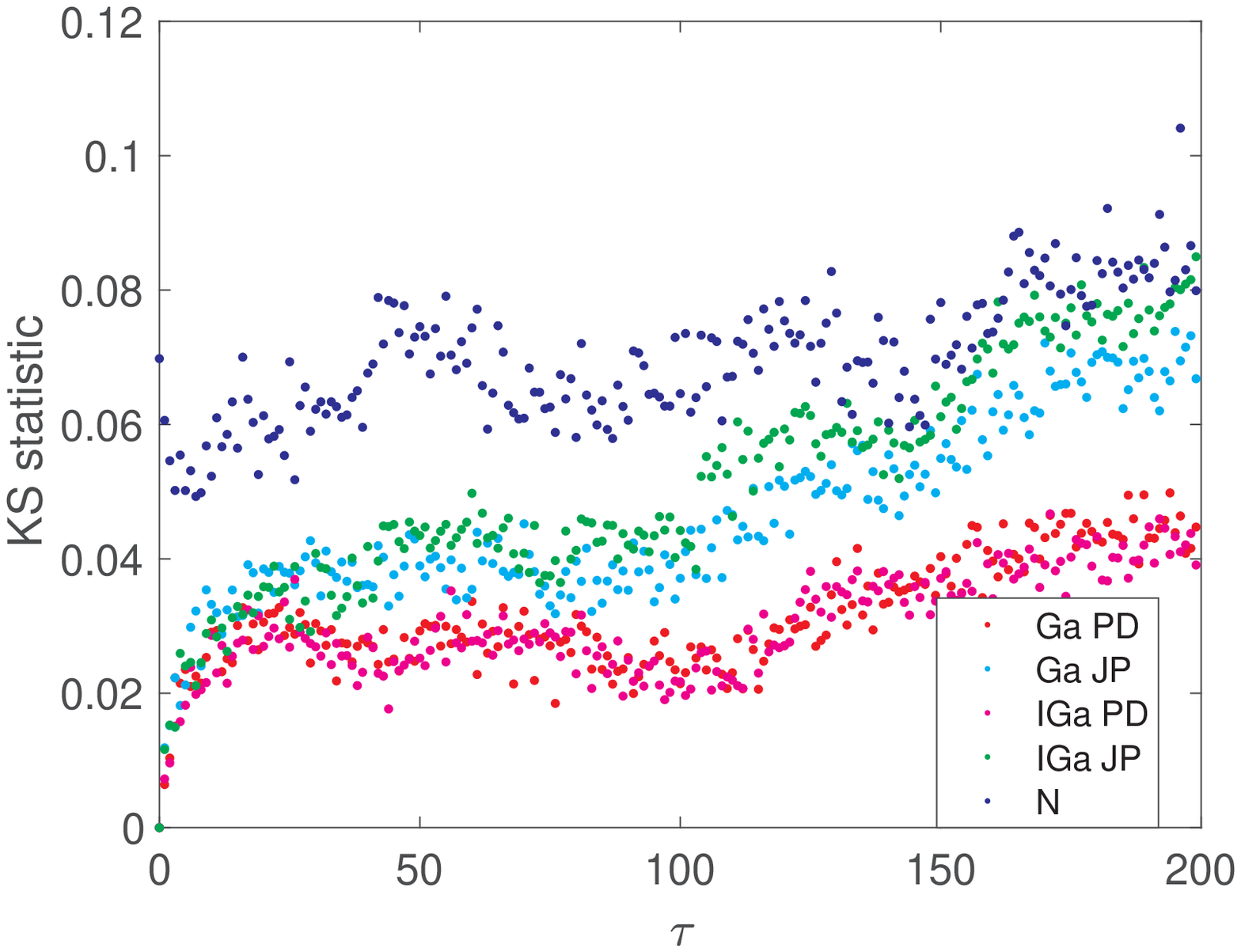}
\end{tabular}

\caption{KS test for DJIA (left) and S\&P500 (right); ``Ga'' -- Heston model, ``IGa'' -- multiplicative model.}
\label{KStest}
\end{figure}

We perform additional statistical testing using the ratio of log likelihood (LL) of (\ref{pdfH}), (\ref{pdfJH}), (\ref{pdfM}) and (\ref{pdfJM}) to that of N and by evaluating Kolmogorov-Smirnov (KS) statistic. The results are shown in Fig. \ref{LLratio}--\ref{KStest}. Clearly, these do not resolve which model, Heston or multiplicative, is better. Fitting the tails of the distribution  \cite{ma2014model} is inconclusive as well. Examination of the moments of the distribution in the next Section resolves the issue in favor of the HM.

\section{Moments of Stock Returns Distribution\label{Moments}}

Moments of the distributions (\ref{pdfH}) and (\ref{pdfM}) can be easily evaluated. The first three non-zero moments of the former and the first two of the latter are given by (\ref{EH2})--(\ref{EM4}). Notice, that for the MM higher moments may not exist due to power-law tails of the Student's distribution (\ref{pdfM}). For instance, $E_M(z^6)$ exist only for $\alpha > 2 \theta$, which is not the case here as is obvious from Fig. \ref{UYSPIAGoogleAllalphaList}. 

\begin{equation}
E_H(z^2) = \theta \tau
\label{EH2}
\end{equation}
\begin{equation}
E_H(z^4) = \frac{3(1 + \alpha)\theta^2\tau^2}{\alpha}
\label{EH4}
\end{equation}
\begin{equation}
E_H(z^6) = \frac{15(1 + \alpha)(2 + \alpha)\theta^3\tau^3}{\alpha^2}
\label{EH6}
\end{equation}
%\begin{equation}
%E_H(z^8) = \frac{105(1 + \alpha)(2 + \alpha)(3 + \alpha)\theta^4\tau^4}{\alpha^3}
%\label{EH8}
%\end{equation}
%\begin{equation}
%E_H(z^{10}) = \frac{945(1 + \alpha)(2 + \alpha)(3 + \alpha)(4 + \alpha)\theta^5\tau^5}{\alpha^4}
%\label{EH10}
%\end{equation}
%\begin{equation}
%E_H(z^{12}) = \frac{10395(1 + \alpha)(2 + \alpha)(3 + \alpha)(4 + \alpha)(5 + \alpha)\theta^6\tau^6}{\alpha^5}
%\label{EH12}
%\end{equation}

\begin{figure}[!htbp]
\centering
\begin{tabular}{cc}
\includegraphics[width = 0.4 \textwidth]{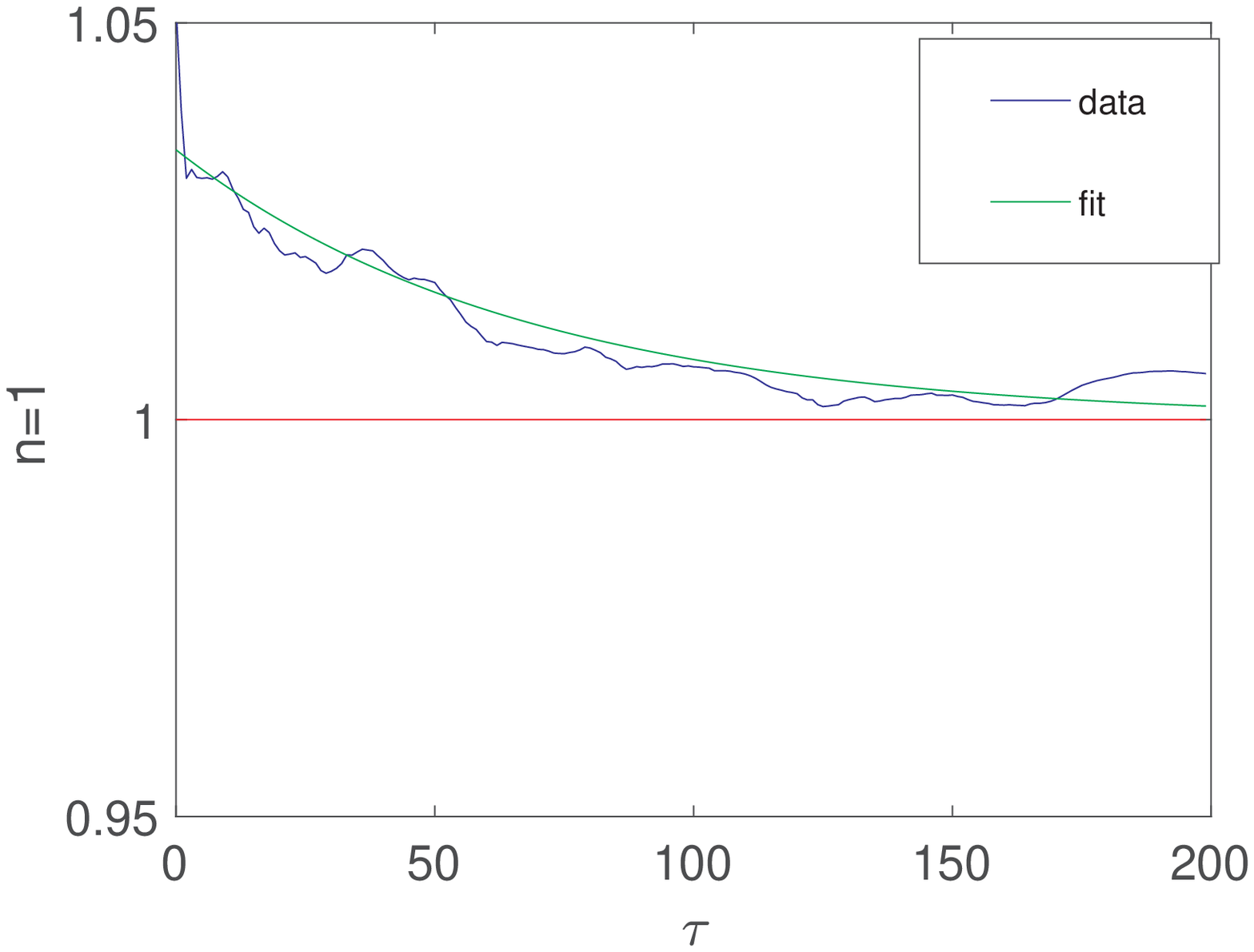}
\includegraphics[width = 0.4 \textwidth]{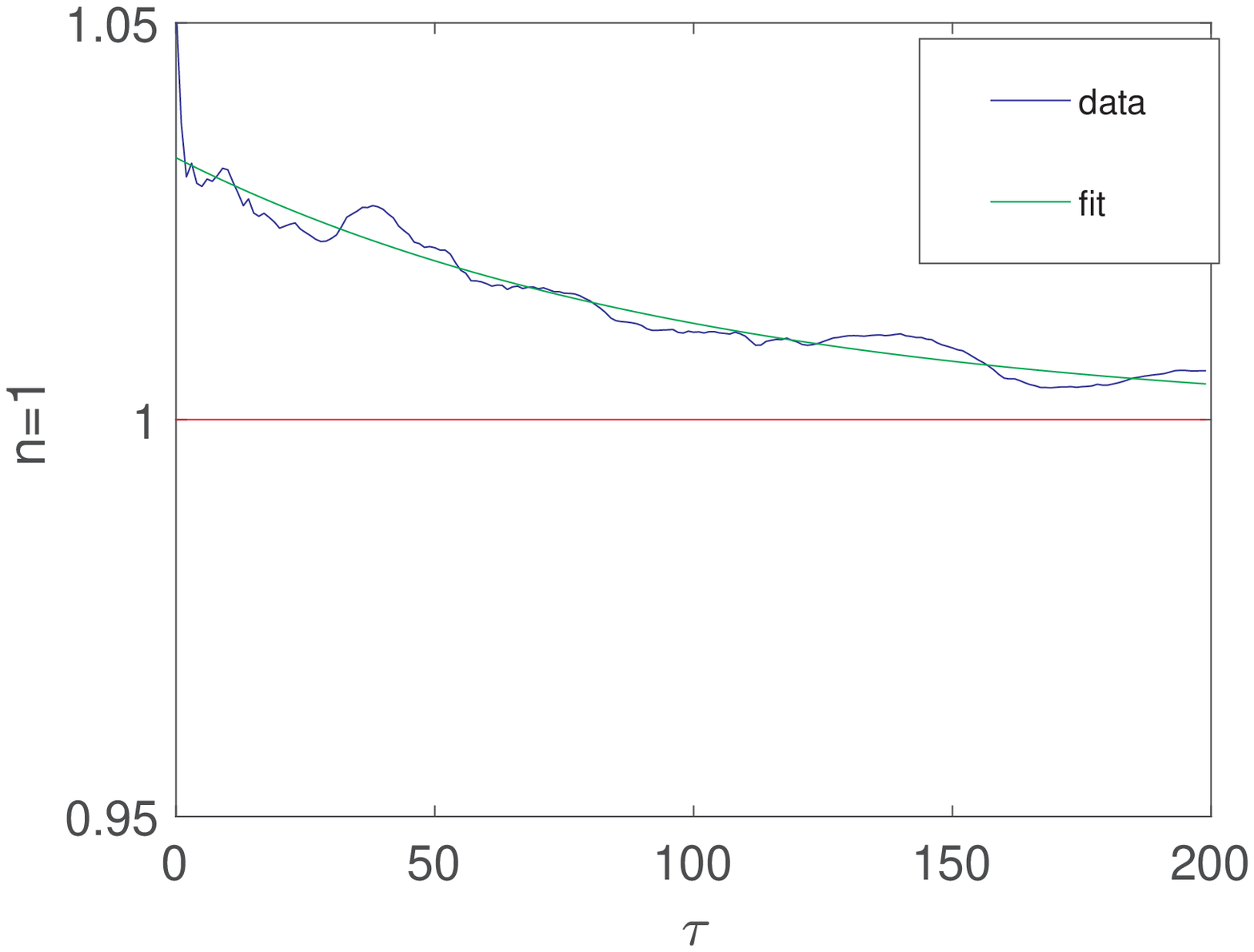}
\end{tabular}
\begin{tabular}{cc}
\includegraphics[width = 0.4 \textwidth]{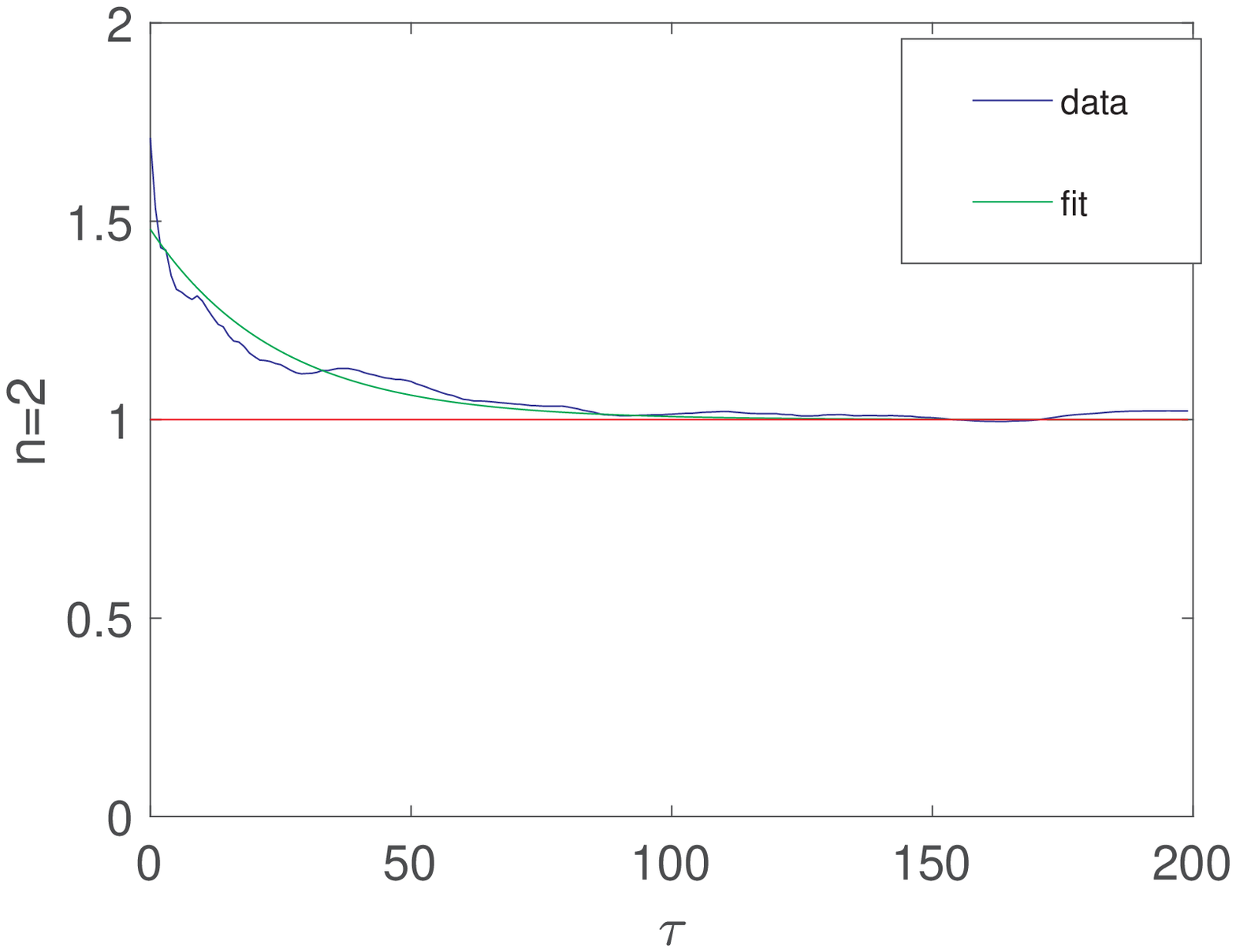}
\includegraphics[width = 0.4 \textwidth]{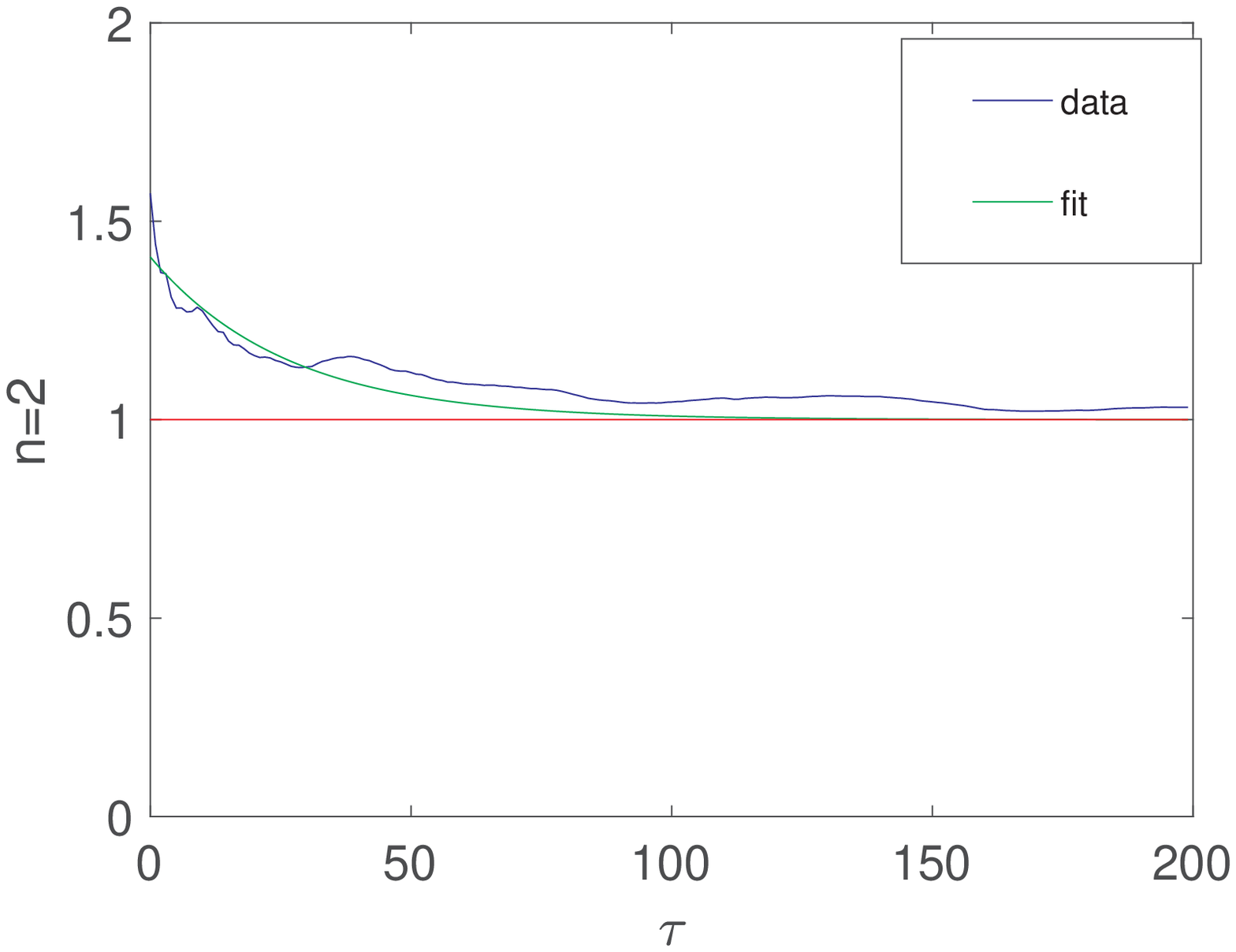}
\end{tabular}
\begin{tabular}{cc}
\includegraphics[width = 0.4 \textwidth]{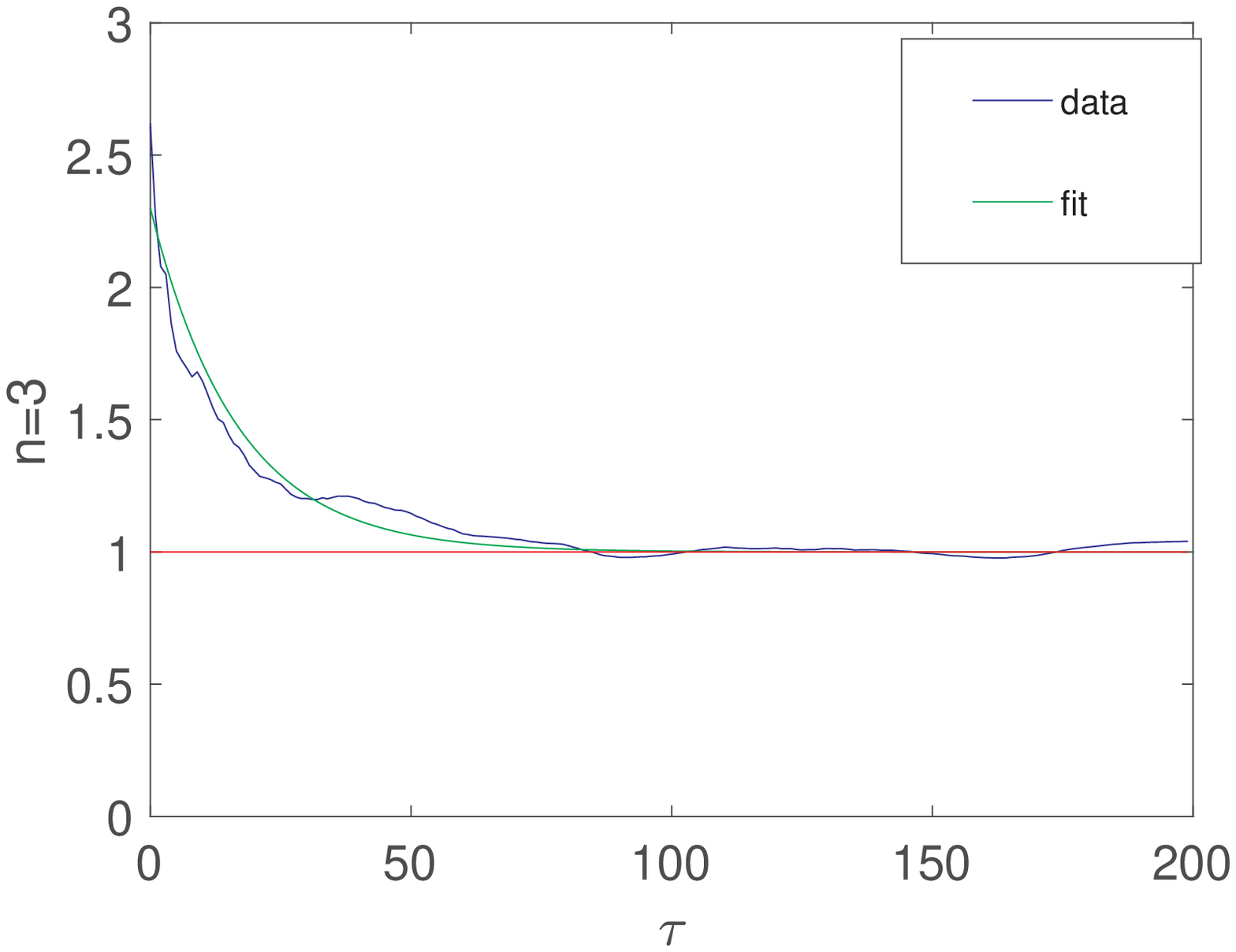}
\includegraphics[width = 0.4 \textwidth]{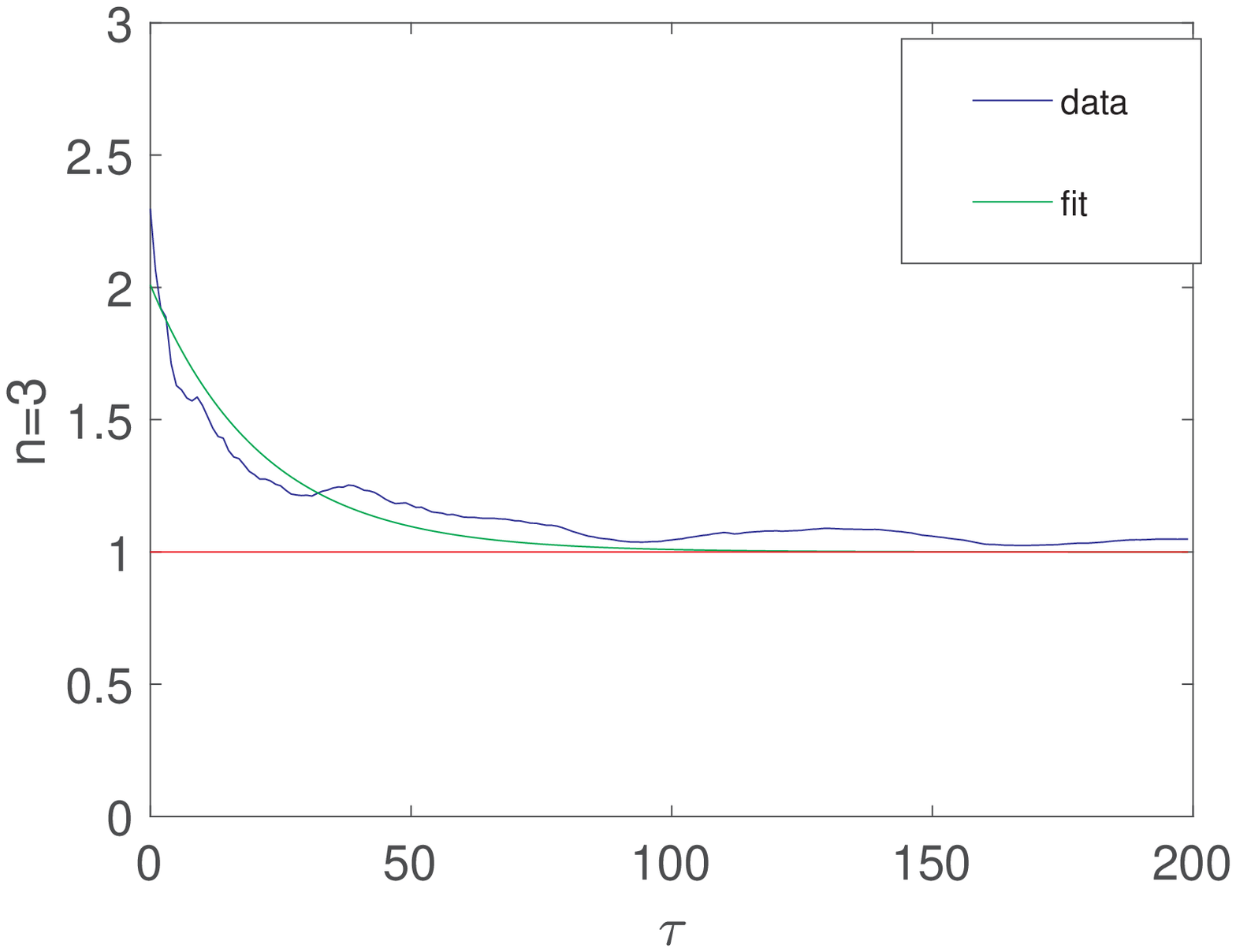}
\end{tabular}

\caption{Heston model, (\ref{EH2})--(\ref{EH6}): From top to bottom, $\left(\frac{\overline{z^{2n}}}{E_H(z^{2n})}\right)^{\frac{1}{2 n}}$ for $n=1$, 2 and 3 respectively, fitted with $1 + b \exp{(-a \tau)}$, for DJIA (left) and S\&P500 (right).}
\label{Gacumulantsfit246}
\end{figure}

\begin{figure}[!htbp]
\centering
\begin{tabular}{cc}
\includegraphics[width = 0.4 \textwidth]{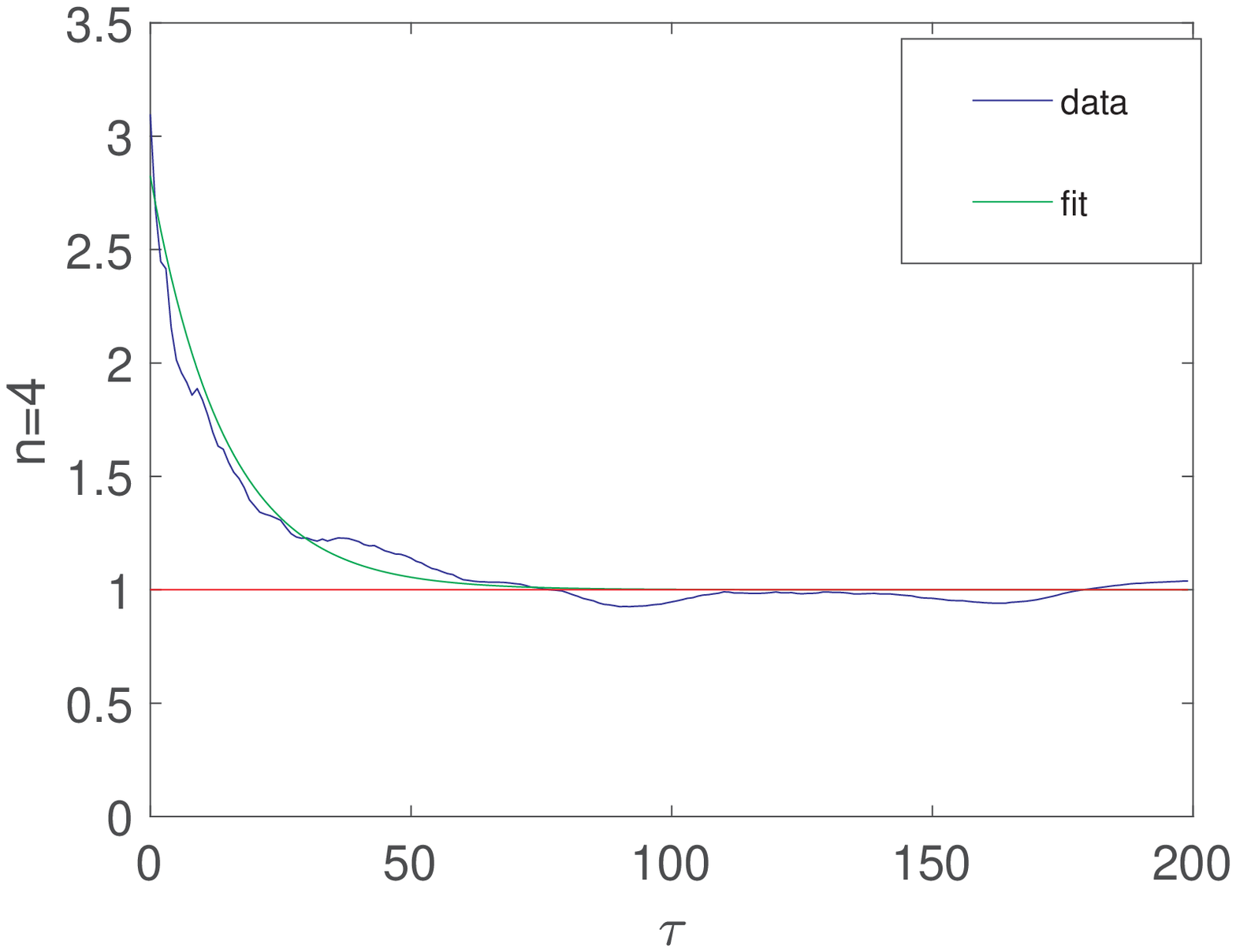}
\includegraphics[width = 0.4 \textwidth]{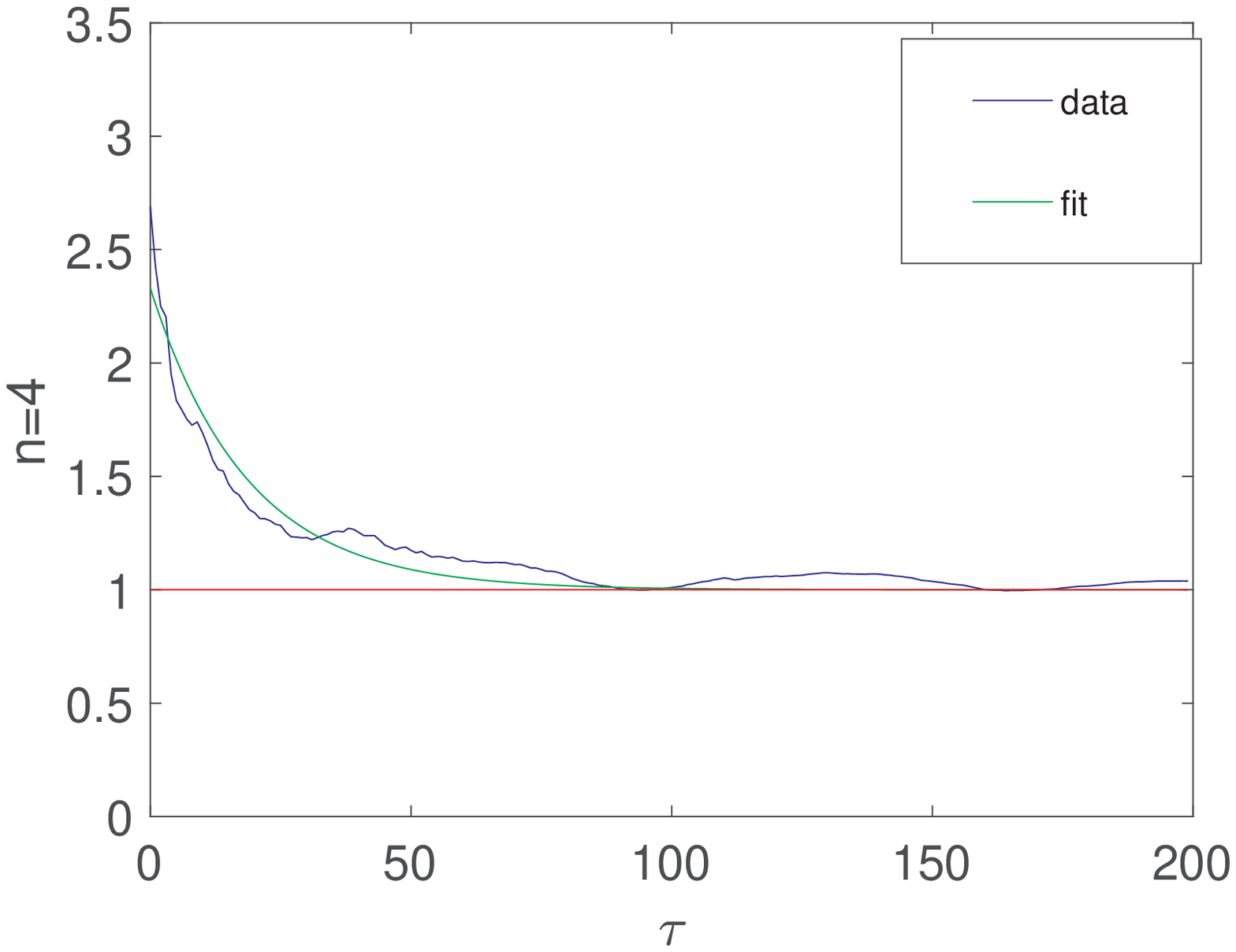}
\end{tabular}
\begin{tabular}{cc}
\includegraphics[width = 0.4 \textwidth]{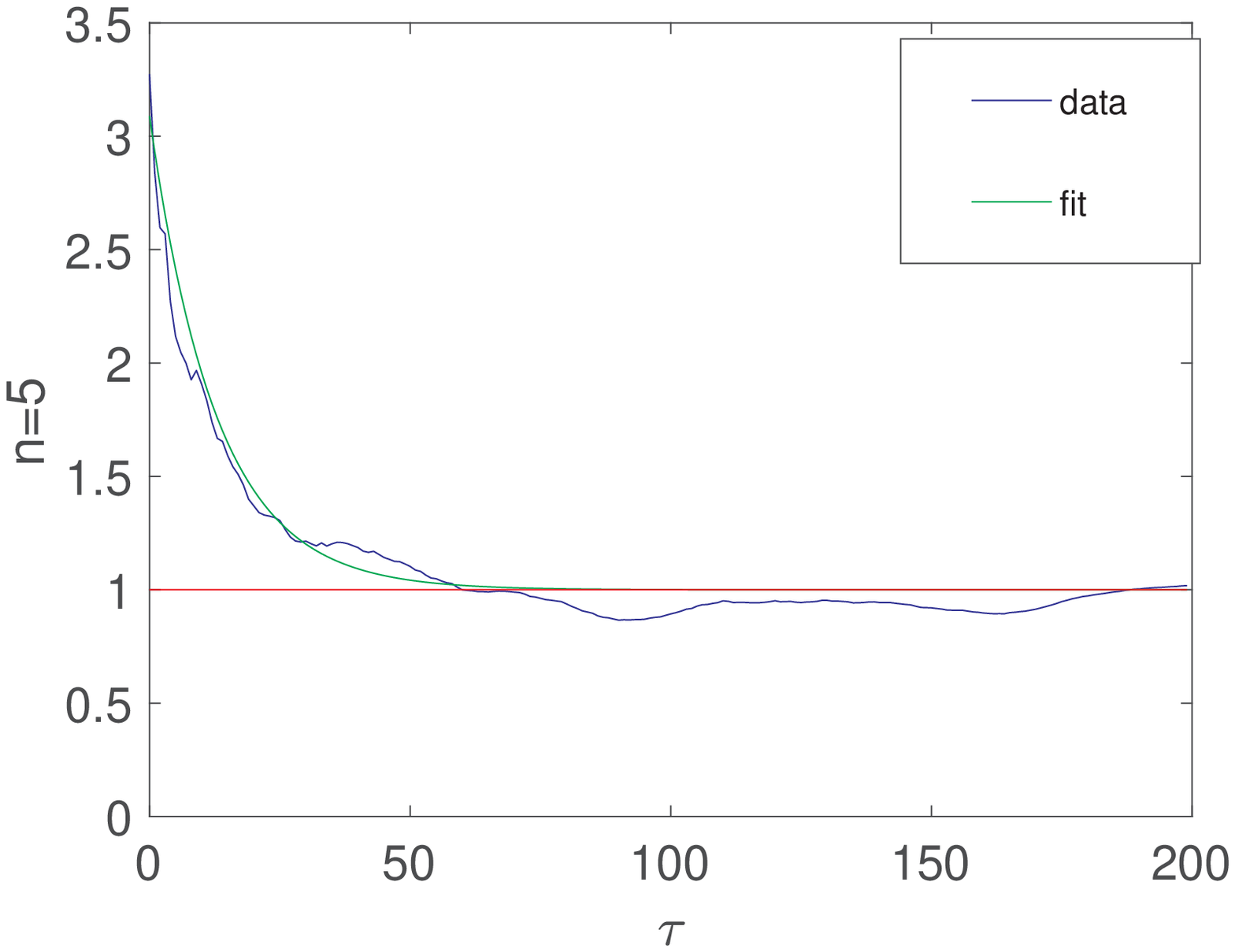}
\includegraphics[width = 0.4 \textwidth]{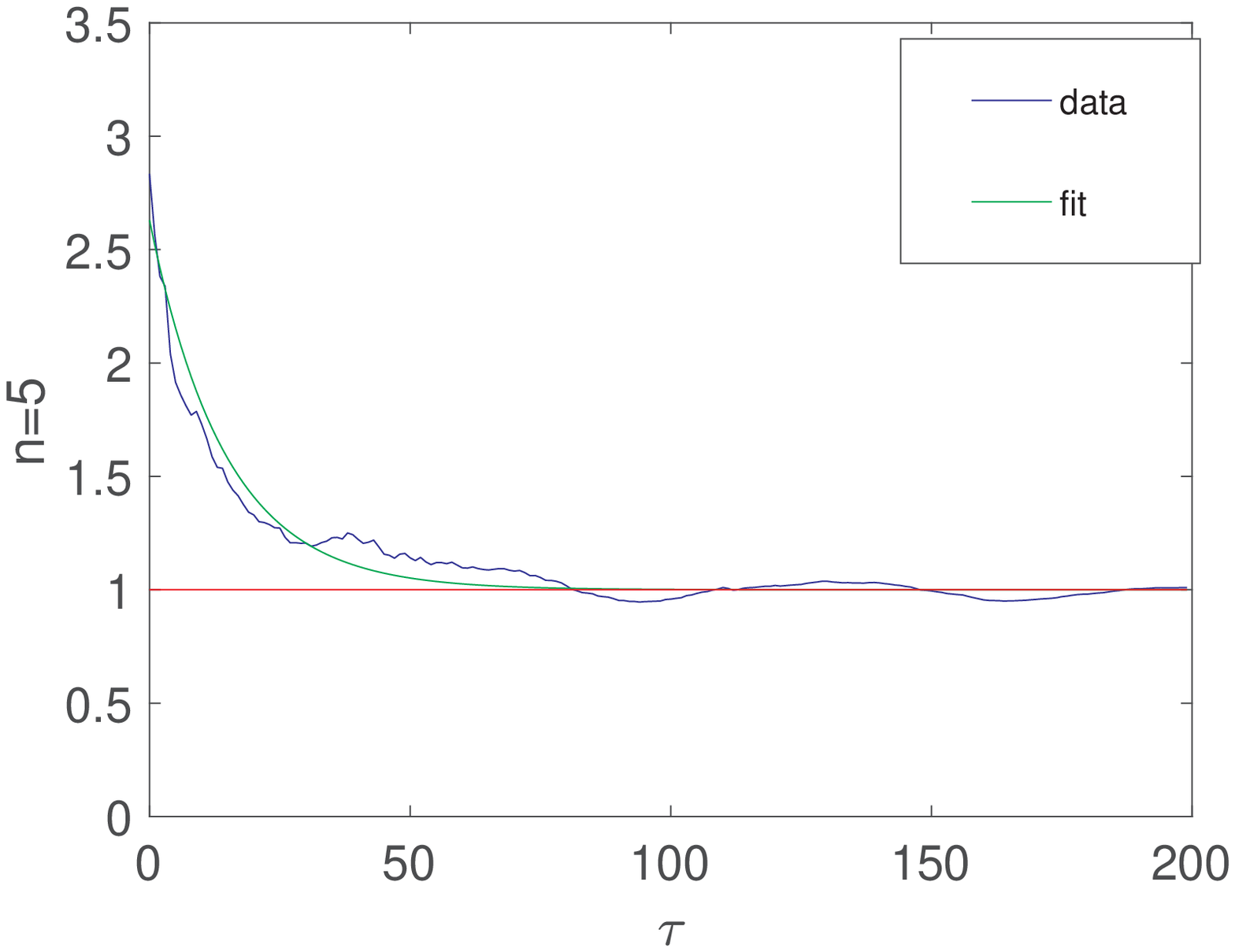}
\end{tabular}
\begin{tabular}{cc}
\includegraphics[width = 0.4 \textwidth]{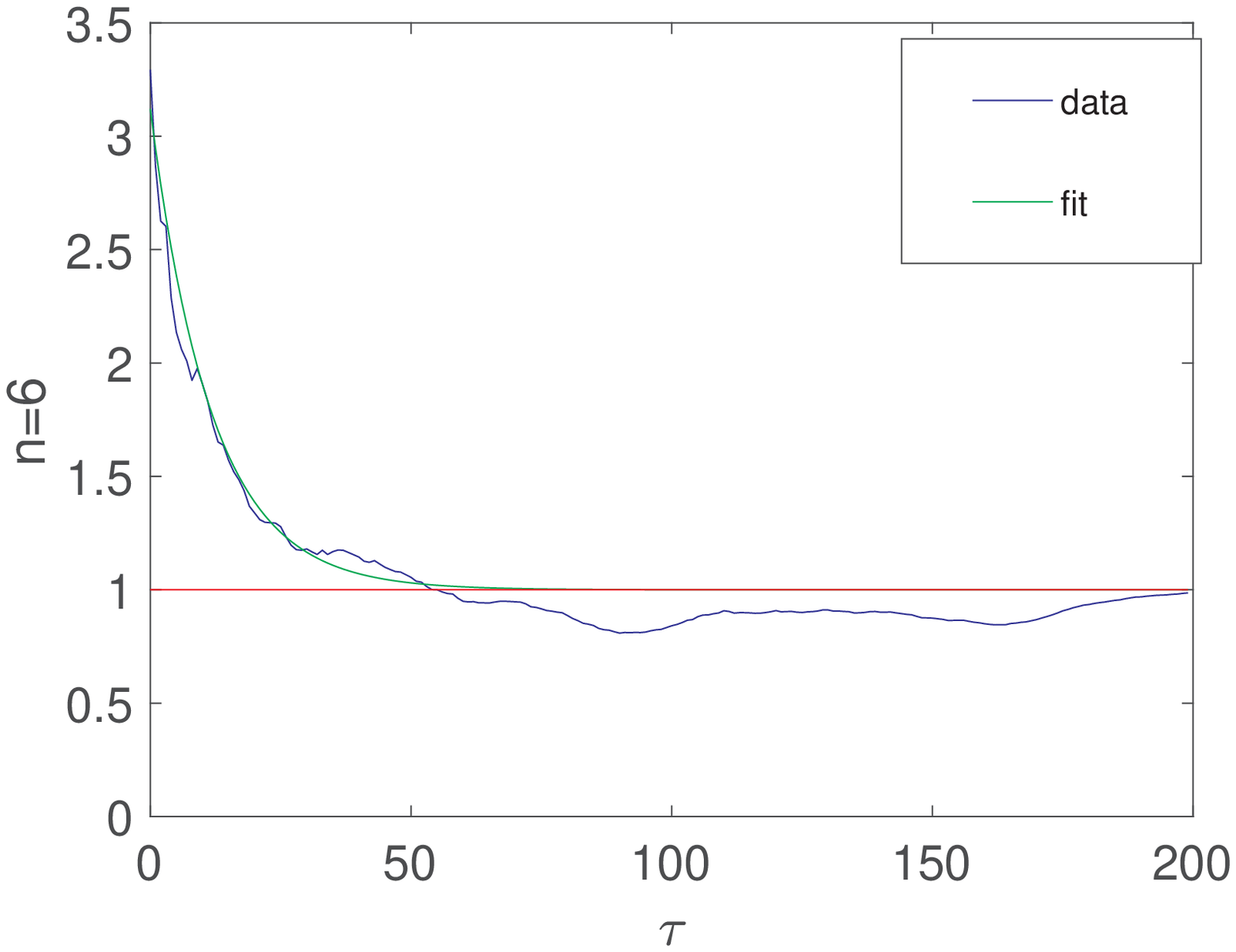}
\includegraphics[width = 0.4 \textwidth]{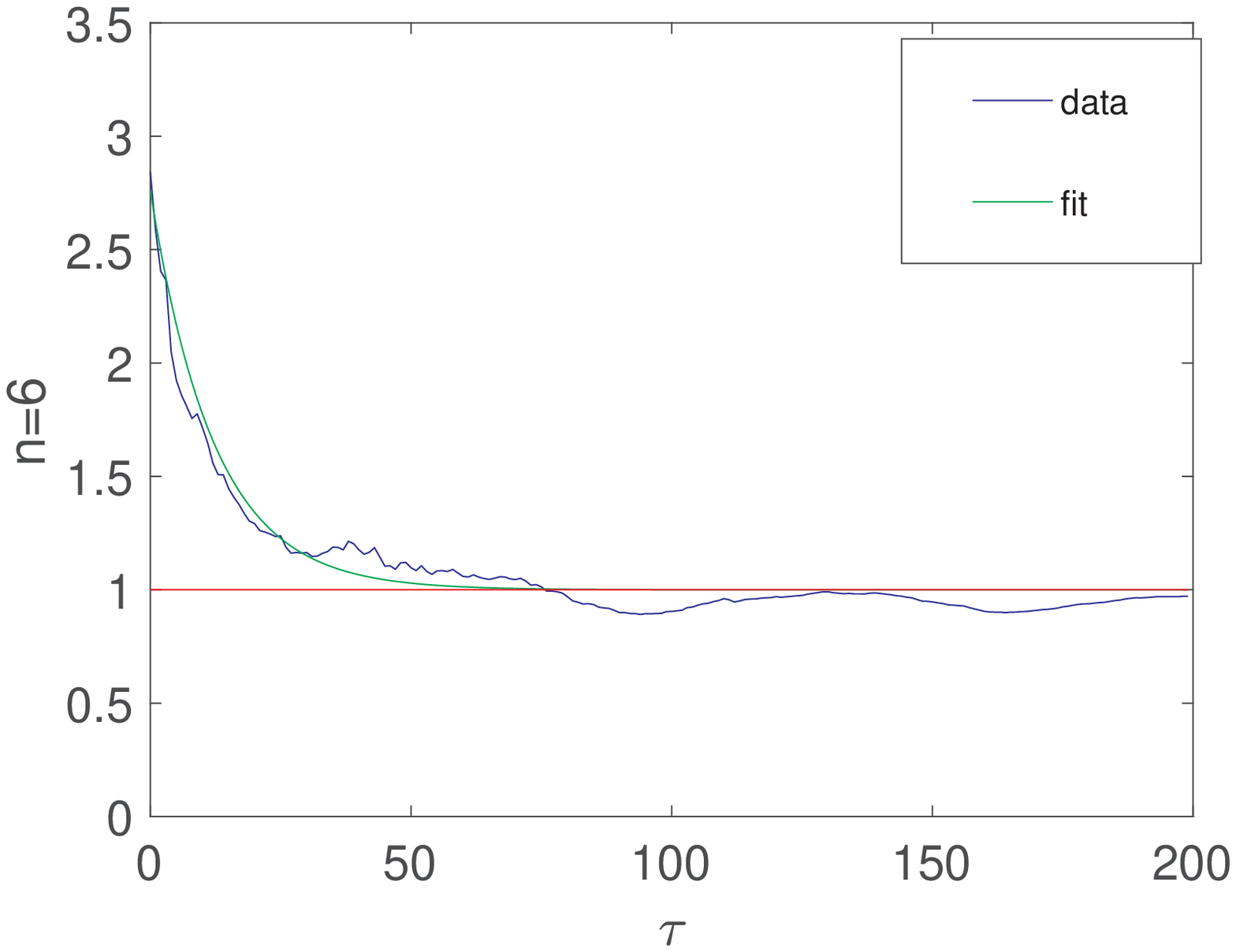}
\end{tabular}

\caption{Heston model: From top to bottom, $\left(\frac{\overline{z^{2n}}}{E_H(z^{2n})}\right)^{\frac{1}{2 n}}$ for $n=4$, 5 and 6 respectively, fitted with $1 + b \exp{(-a \tau)}$, for DJIA (left) and S\&P500 (right).}
\label{Gacumulantsfit81012}
\end{figure}

\begin{equation}
E_M(z^2) = \theta \tau
\label{EM2}
\end{equation}
\begin{equation}
E_M(z^4) = \frac{3\alpha\theta^2\tau^2}{(\alpha - \theta)}
\label{EM4}
\end{equation}

\begin{figure}[!htbp]
\centering
\begin{tabular}{cc}
\includegraphics[width = 0.4 \textwidth]{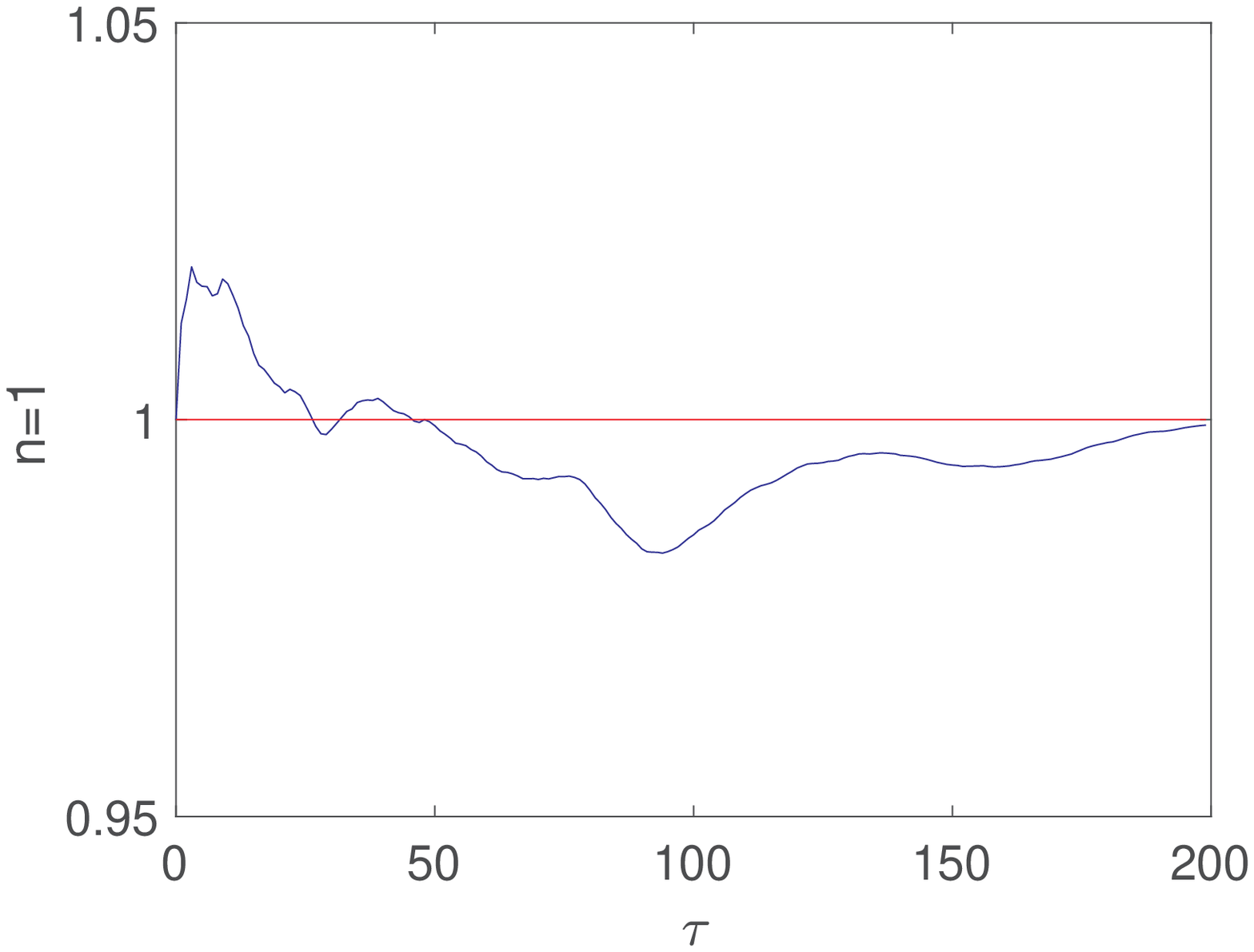}
\includegraphics[width = 0.4 \textwidth]{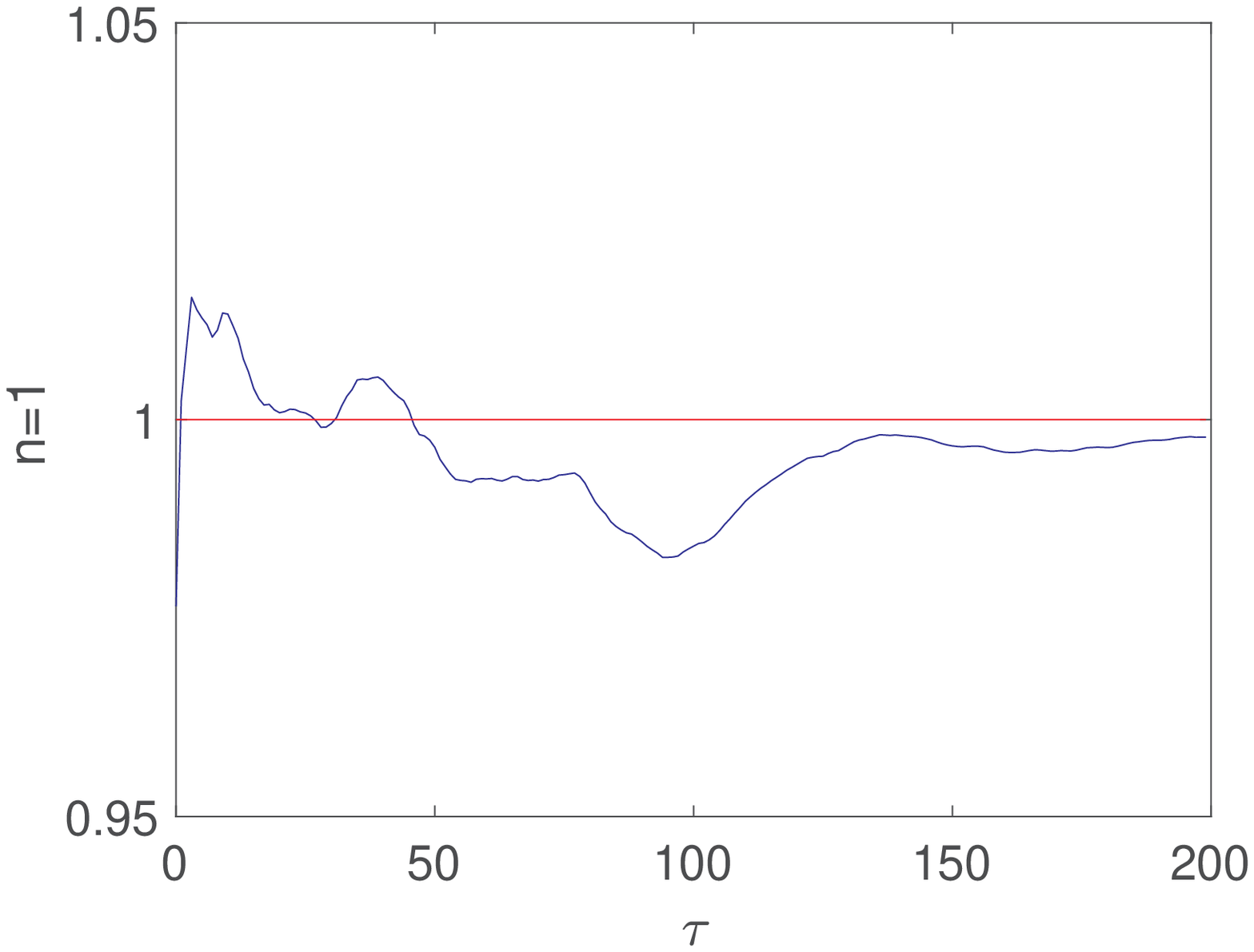}
\end{tabular}
\begin{tabular}{cc}
\includegraphics[width = 0.4 \textwidth]{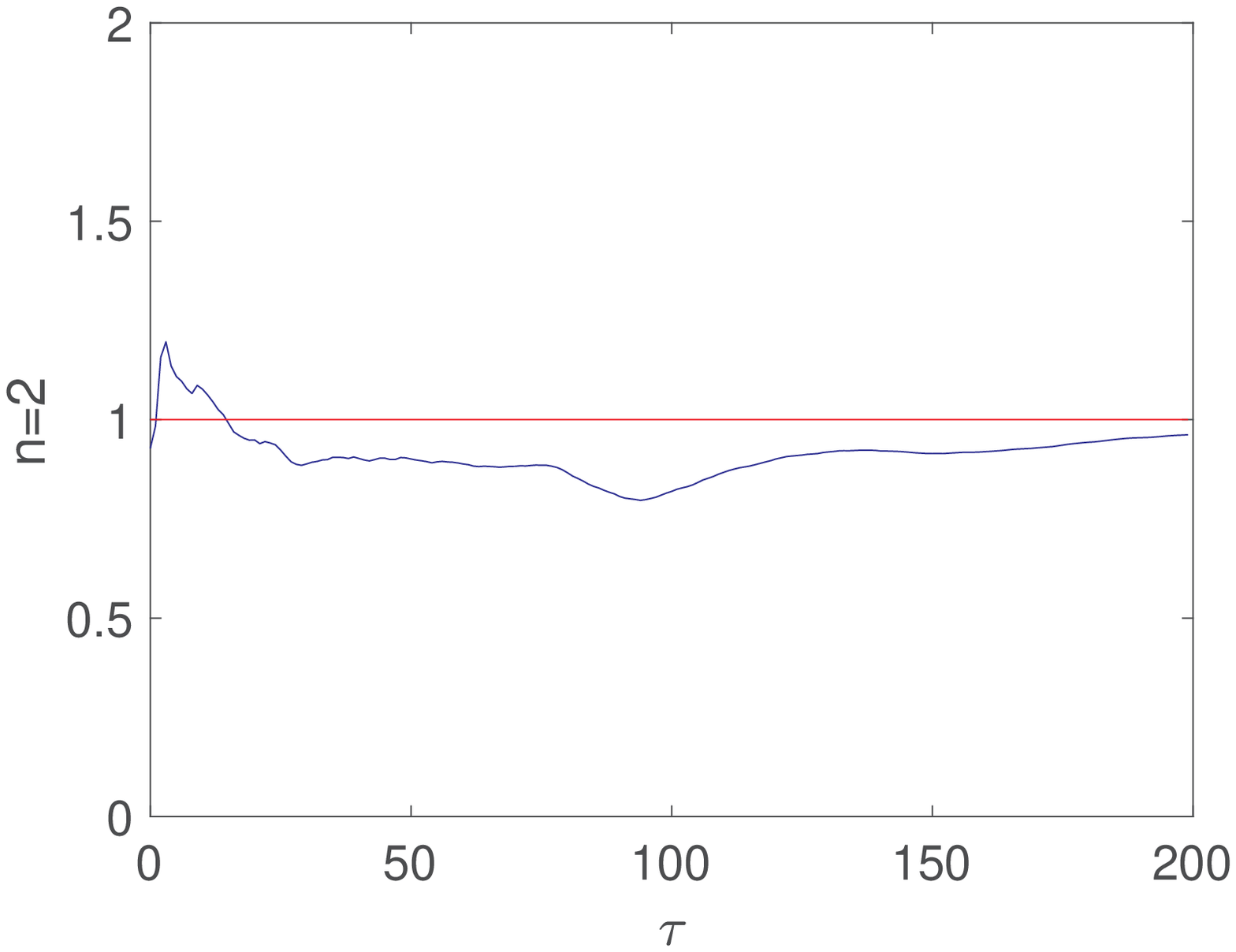}
\includegraphics[width = 0.4 \textwidth]{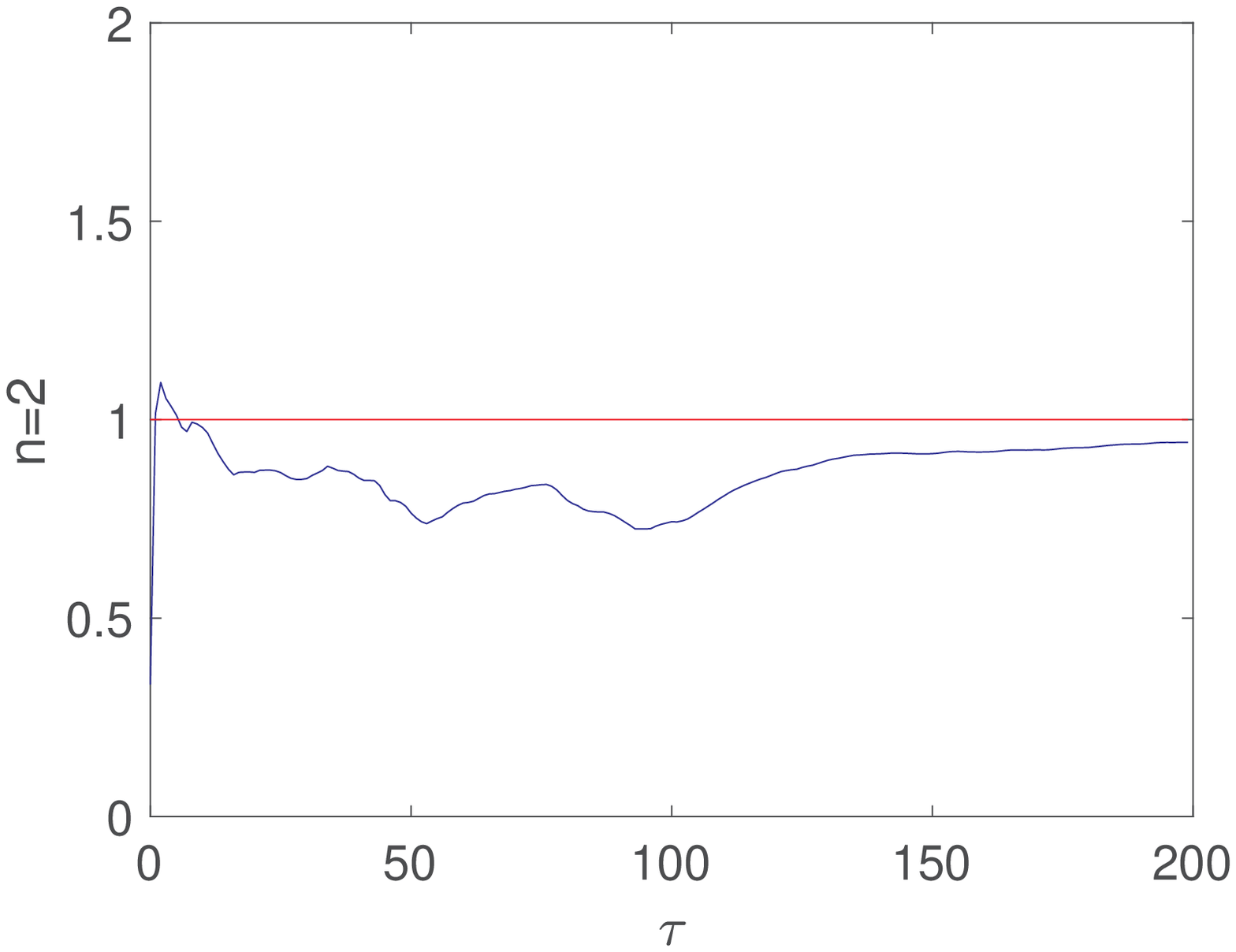}
\end{tabular}
\caption{Multiplicative model, (\ref{EM2})--(\ref{EM4}): From top to bottom, $\left(\frac{\overline{z^{2n}}}{E_M(z^{2n})}\right)^{\frac{1}{2 n}}$ for $n=1$ and 2 respectively for DJIA (left) and S\&P500 (right).}
\label{IGa Cumulants}
\end{figure}

The moments $\overline{z^2}$, $\overline{z^4}$, and $\overline{z^6}$ obtained from the DJIA and S\&P500 data are compared with the theoretical expressions by plotting $\left(\frac{\overline{z^{2n}}}{E(z^{2n})}\right)^{\frac{1}{2 n}}$. Figs. \ref{Gacumulantsfit246}--\ref{IGa Cumulants} clearly show excellent agreement with the HM. The approach to unity is fitted for the latter with the exponential relaxation $1 + b \exp{(-a \tau)}$ and the results for the parameters are collected in Table \ref{fitted result}. The value of $b$, which describes relaxation of the moments, correlates well with the value of $\gamma \approx 0.045-0.05$, which can be obtained from leverage  \cite{bouchaud2001leverage,perello2002correlated,perello2004multiple}  and is somewhat model-dependent.

\begin{table}[!htbp]
\centering
\begin{tabular}{ccccc} 
\hline
			&      			   DJIA 	&    & S\&P500 \\
\hline
           	&          	a&			b&         		a&				b\\
\hline
 n=1 		&		0.015 &           0.034 &		0.010 &			0.033\\
\hline
 n=2		&		0.041 & 		0.48 &		0.038 &			0.41\\
\hline
 n=3 		&		0.060 &		1.30 &		0.047 & 			1.01  \\
\hline
 n=4		&		0.070 &		1.82 &		0.054 &			1.33 \\
\hline
n=5		&		0.078 &		2.09 &		0.069& 			1.63 \\
\hline
n=6		&           0.0085 & 		2.12 & 		0.082 & 			1.76\\
\hline
\end{tabular}
\caption{Fitting parameters in Figs. \ref{Gacumulantsfit246}--\ref{Gacumulantsfit81012} }
\label{fitted result}
\end{table}

\newpage

\section{Conclusions\label{Conclusions}}
Our main conclusions can be summarized as follows:
\begin{itemize}
  \item In the mean-reverting models of volatility, the mean value determines the mean realized variance, which scales linearly with the number of days over which the returns are accumulated, as per (\ref{RVcalc}). It also, by definition, determines the width of the distribution function of stock returns. For Heston and multiplicative models, (\ref{sdrGaGa}) and (\ref{sdrIGaIGa}), this is illustrated by (\ref{EH2}) and (\ref{EM2}) and Fig. \ref{Probability Distribution of Stock Returns}.
  \item Over long periods of accumulation of stock returns -- measured in years -- the distribution of stock returns can be found as a product distribution of the steady-state distribution of volatility and normal distribution representing Wiener noise. For Heston and multiplicative models these are given, respectively, by (\ref{pdfH}) and (\ref{pdfM}). Exact forms can be also derived, (\ref{pdfJH}) and (\ref{pdfJM}), but they give no discernible differences over periods of interest.
    \item Analysis of the moments of the distribution indicate that the Heston model is more suitable than the multiplicative model.
\end{itemize}
\appendix
\section{}

\subsection{Joint PDF Derivation}

To take into account the Ito term in Eq. (\ref{sdrxt}), we denote $Z = \mathrm{d}x_t$, $X = - \frac{v_t}{2}\mathrm{d}t$ and $Y = \sqrt{v_t}\mathrm{d}W_t^{(1)}$. Since $X$ and $Y$ are dependent variables, we need the joint PDF of $X$ and $Y$ in order to obtain the distribution of $Z$:
\begin{equation}
f_{X,\, Y}(x, y) = f_{Y|X}(y|x) f_{X}(x)
\end{equation}
Since $X$ is aproduct of a constant $-\frac{\mathrm{d}t}{2}$ with $v_t$, we immediately obtain $f_{X}(x)$ as
\begin{equation}
f_{X}(x) = \frac{2}{\mathrm{d}t} \alpha \mathrm{Ga(} -\alpha \frac{2x}{\mathrm{d}t} ;\, \alpha,\, \theta\mathrm{)}
\label{Happ}
\end{equation}
for the HM and 
\begin{equation}
f_{X}(x) = \frac{2}{\mathrm{d}t} \mathrm{IGa(} -\frac{2x}{\mathrm{d}t} ;\, \frac{\alpha + \theta}{\theta},\, \alpha\mathrm{)}
\label{Mapp}
\end{equation}
for the MM. On the hand, $f_{Y|X}(y|x)$ can be obtained as follows:
\begin{equation}
f_{Y|X}(y|x = -\frac{v_t}{2} \mathrm{d}t) = \frac{ \mathrm{N(} \frac{y}{\sqrt{-\frac{2x}{\mathrm{d}t}}};\, 0, \mathrm{d}t \mathrm{)} }
{ \int_{-\infty}^{+\infty} \mathrm{N(} \frac{\varepsilon}{\sqrt{-\frac{2x}{\mathrm{d}t}}};\, 0, \mathrm{d}t \mathrm{)} \mathrm{d}\varepsilon }
= \frac{\sqrt{\mathrm{d}t}}{\sqrt{-2x}} \mathrm{N(} \frac{y}{\sqrt{-\frac{2x}{\mathrm{d}t}}};\, 0, \mathrm{d}t \mathrm{)}
\end{equation}
whereof the distribution of $Z$ is found as
\begin{equation}
\phi(z) = \int_{-\infty}^{0} f_{X,\,Y}(x,\, z-x) \mathrm{d}x 
\end{equation}
Integrating and replacing $\mathrm{d}t$ with $\tau$, we obtain (\ref{pdfJH}) for (\ref{Happ}) and (\ref{pdfJM}) for (\ref{Mapp}).

\subsection{Product Distribution vs. Joint Distribution for the Multiplicative Model}

The most direct comparison of JP to PD is accomplished by taking the ratio of (\ref{pdfJM}) to (\ref{pdfM}) and comparing it to unity. Operating under condition $\theta \tau \ll 1$ is roughly the same as $\alpha \tau \ll 1$ since $\theta \sim \alpha$ for MM. expanding in $z$ and in $\alpha \tau$, and keeping the two leading term for the constant and $z$, yields
\begin{equation}
\frac{\psi_M(z)}{\phi_M(z)} \approx 1- \frac{z}{2} +\frac{(-2+z) \theta \alpha \tau}{8 (2\alpha+\theta)} 
\end{equation}
Obviously, the corrections to unity are very small under the assumptions used. Together with derivation in Sec. \ref{DistrSR} for HM, this now fully explains why JP and PD are hardly distinguishable in Sec. \ref{Numerics}.

\bibliography{mybib}

\end{document}